\documentclass[sigconf]{acmart}
\AtBeginDocument{%
  }

\usepackage[utf8]{inputenc}
\usepackage{graphicx} 
\usepackage{comment}
\usepackage{tikz}
\usepackage{graphicx}
\usepackage{xspace}
\usepackage{xcolor}
\usepackage{xstring}
\usepackage{color, colortbl}
\usepackage{array}
\usepackage{multicol,multirow}
\usepackage{enumitem}
\usepackage{booktabs}
\usepackage{caption}
\usepackage{subcaption}
\usepackage{fontawesome}
\usepackage{soul}
\usepackage{xurl}
\usepackage{listings}
\usepackage{balance}
\usepackage{makecell}
\usepackage{amsfonts}
\usepackage{amsmath}
\usepackage{microtype}
\usepackage{nicematrix}
\usepackage{relsize}
\usepackage[bookmarksnumbered,unicode,hidelinks]{hyperref}
\usepackage{enumitem}
\usepackage[english]{babel}
\usepackage{amsthm}
\usepackage{pifont}

\usepackage[framemethod=tikz]{mdframed}

\usepackage{longtable}

\usepackage{wasysym}

\usepackage{tabularx}
\DeclareRobustCommand{\DashCircle}{%
  \tikz[baseline=-0.6ex, line width=0.4pt]{
    \foreach \a in {0,45,...,315}
      \draw (\a:0.75ex) arc (\a:\a+28:0.75ex);
  }}

\newcommand{\etal}{\textit{et al}.\xspace}

\newcommand{\eg}{\textit{e}.\textit{g}.}

\newif\ifdraft\drafttrue
\newif\ifnotes\notestrue
\ifdraft\else\notesfalse\fi
\input{glyphtounicode}
\newcommand{\squishlist}{
\begin{itemize}[noitemsep,nolistsep]
  \setlength{\itemsep}{-0pt}
}
\newcommand{\squishend}{
  \end{itemize}
}

\newcommand{\PP}[1]{
\noindent{\bf \IfEndWith{#1}{.}{#1}{#1.}}
}
\newcommand{\PPNS}[1]{
\noindent{\bf \IfEndWith{#1}{.}{#1}{#1.}}
}

\newcommand{\boxbeg}{
\vspace{2px}
\noindent\begin{tabular}{|l|}\hline
\begin{minipage}{3.2in}
\vspace{2px}
\noindent
}
\newcommand{\boxend}{
\vspace{2px}
\end{minipage}\\ \hline
\end{tabular}
\vspace{-10pt}
}

\everypar{\looseness=-1}
\definecolor{darkpastelgreen}{rgb}{0.01, 0.75, 0.24}
\definecolor{cadmiumgreen}{rgb}{0.0, 0.42, 0.24}
\definecolor{brickred}{rgb}{0.8, 0.25, 0.33}
\definecolor{cornellred}{rgb}{0.7, 0.11, 0.11}
\definecolor{burgundy}{rgb}{0.5, 0.0, 0.13}
\definecolor{frenchblue}{rgb}{0.0, 0.45, 0.73}
\definecolor{light-gray}{gray}{0.92}
\definecolor{lightlight-gray}{gray}{0.97}
\definecolor{codegray}{gray}{0.90}
\definecolor{inputgray}{gray}{0.90}
\definecolor{dkgreen}{rgb}{0,0.6,0}
\definecolor{gray}{rgb}{0.5,0.5,0.5}
\definecolor{mauve}{rgb}{0.58,0,0.82}
\definecolor{apricot}{rgb}{0.98, 0.81, 0.69}
\definecolor{bubblegum}{rgb}{0.99, 0.76, 0.8}
\global\mdfdefinestyle{rtboxstyle}{
    backgroundcolor=black!4,
    linecolor=black!65,
    linewidth=0.4pt,
    roundcorner=0pt,
    leftmargin=0pt,
    rightmargin=0pt,
    innerleftmargin=5pt,
    innerrightmargin=5pt,
    innertopmargin=4pt,
    innerbottommargin=4pt,
    skipabove=6pt,
    skipbelow=6pt,
    nobreak=true
}

\DeclareRobustCommand*\BC[1]{%
\begin{tikzpicture}[baseline=(C.base)]
\node[draw,circle,fill=black,inner sep=0.2pt](C) {\textcolor{white}{#1}};
\end{tikzpicture}}

\usepackage{tikz}
\usetikzlibrary{patterns}
\usepackage{xspace}
\definecolor{tlsred}{HTML}{E5372B}    
\definecolor{tlsgreen}{HTML}{3A8F3A}

\DeclareRobustCommand{\tlsclient}{%
  \tikz[baseline=-0.5ex]\draw[draw=red!60!black, fill=red!85!black,
        line width=0.3pt, rounded corners=0.4pt]
        (-0.8ex,-0.8ex) rectangle (0.8ex,0.8ex);}

\DeclareRobustCommand{\tlsserver}{%
  \tikz[baseline=-0.5ex]\draw[draw=green!55!black, line width=0.3pt,
        pattern=north east lines, pattern color=green!55!black]
        (-0.8ex,-0.8ex) rectangle (0.8ex,0.8ex);}

\everypar{\looseness=-1}

\usepackage{nameref}

\DeclareRobustCommand*\BC[1]{%
\begin{tikzpicture}[baseline=(C.base)]
\node[draw,circle,fill=black,inner sep=0.2pt](C) {\textcolor{white}{#1}};
\end{tikzpicture}}
\usepackage[most]{tcolorbox}

\newtcolorbox{boxI}{
    colback = lightgray!10, 
    colframe = black, 
    boxrule = 0.5pt, 
    toprule = 0.5pt, 
    arc = 2pt,
    left = 1pt,
    right = 1pt,
    bottom = 0pt,
    top = 0pt
}
\newcounter{observcntr}

\renewcommand\footnotetextcopyrightpermission[1]{} 
\setcopyright{none}

\begin{document}
\author{Jaehwan Park}
\email{jpark127@utk.edu}
\affiliation{%
  \institution{University of Tennessee}
  \city{Knoxville}
  \country{USA}
}

\author{Sunghyun Park}
\email{bags3366@khu.ac.kr}
\affiliation{%
  \institution{Kyung Hee University}
  \city{Yongin}
  \country{South Korea}
}

\author{Kevin Nam}
\authornote{Corresponding authors.}
\email{kvnam@khu.ac.kr}
\affiliation{%
  \institution{Kyung Hee University}
  \city{Yongin}
  \country{South Korea}
}

\author{Doowon Kim}
\authornotemark[1]
\email{doowon@utk.edu}
\affiliation{%
  \institution{University of Tennessee}
  \city{Knoxville}
  \country{USA}
}

\title{They Will See Johnny's Message: A Measurement Study of Post-Quantum Cryptography Across Email Delivery Paths}


\renewcommand{\shortauthors}{Park et al.}

\begin{abstract}
Harvest-now, decrypt-later (HNDL) attacks make the transition to post-quantum cryptography (PQC) urgent for email.
This paper presents a systematic measurement that jointly examines PQC deployment across email service stages, negotiation by communicating peers, and usable message encryption.
Endpoint-level support alone does not establish PQC readiness across email workflows because submission, relay, and access use independently negotiated Transport Layer Security (TLS) connections.
We therefore combine three snapshots of relay, submission, access, and webmail services associated with the Tranco Top~1M domains with controlled tests of mail clients, sending servers, and OpenPGP and S/MIME implementations.
Our measurements reveal a transition concentrated among a few providers, with available capability not consistently translating into actual use.
Fewer than half of the domains in each service panel support PQC, and Google alone accounts for approximately 87\% of hybrid-capable relay domains.
Only two of eight tested sending providers negotiate PQC with our receiver, although five accept it inbound.
When TLS groups are restricted to PQC, four of the six incompatible senders fall back to plaintext where unencrypted SMTP remains permitted.
At the message layer, limited native key generation, key-format incompatibilities, and backend integration failures further constrain usable PQC encryption.
\end{abstract}


\keywords{Email security, Internet measurement, Post-quantum cryptography}



\maketitle

\section{Introduction}
\label{sec:intro}

\begin{figure*}[t!]
    \centering
    \includegraphics[width=\linewidth]{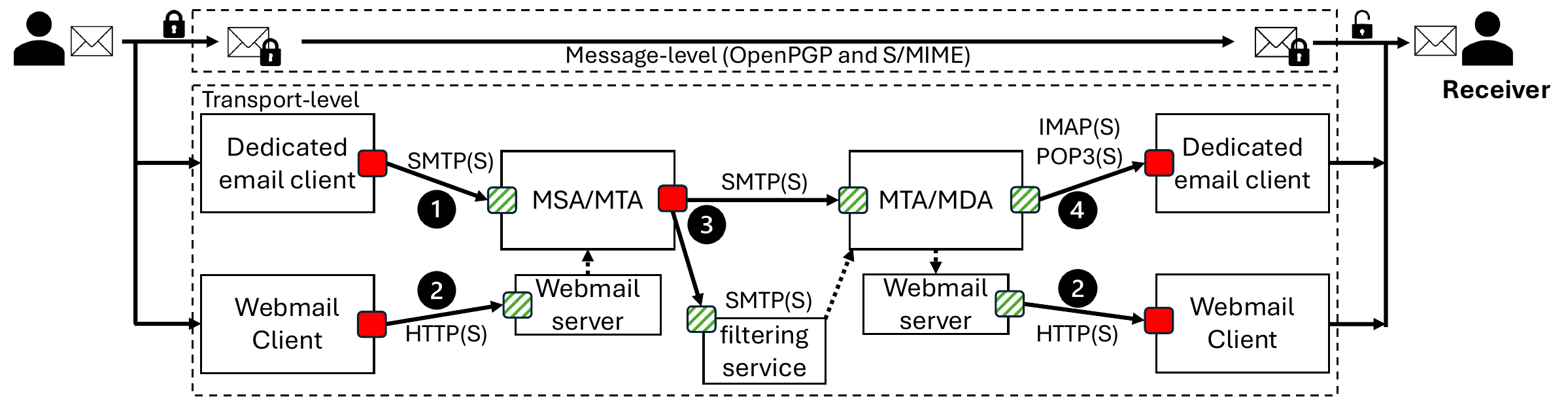}
    \caption{Email delivery and measurement scope.
    The message layer protects content across transport connections.
    The transport layer comprises submission (\ding{182}), webmail (\ding{183}), relay (\ding{184}), and retrieval (\ding{185}).
    \tlsserver\ and \tlsclient\ mark the TLS server and client sides.
    Solid arrows show measured connections.
    Dashed arrows show provider-internal connections that are not measured.}
    \label{fig:email_delivery}
\end{figure*}

Email remains a widely used communication channel~\cite{statista2025emails} carrying sensitive personal and organizational information~\cite{durumeric2015neither}.
Transport Layer Security (TLS) was added to the original protocols and now provides the primary protection for email in transit~\cite{rfc8446,rfc3207,rfc8314}.
Quantum computing threatens this protection.
A cryptographically relevant quantum computer (CRQC) could break traditional TLS key establishment, including RSA and (EC)DHE, and recover the session keys protecting email~\cite{new01,rfc9958}.
The National Institute of Standards and Technology (NIST) standardized ML-KEM to support quantum-resistant key establishment~\cite{nist2024fips203}.

Migration to post-quantum cryptography (PQC) is already urgent because \textit{harvest-now, decrypt-later} (HNDL) attacks allow adversaries to record encrypted traffic today for future decryption~\cite{bhargavan2024formal,cisa2023quantum,sikeridis2020post}.
Email is particularly exposed because correspondence can remain sensitive for years and cross the public Internet between organizations, where an on-path adversary can collect it.
Archived messages could later reveal sensitive information, including political affiliations, when their authors can no longer protect the captured data.
Protecting email against HNDL requires more than securing an individual TLS connection.

Email uses independently negotiated connections for submission, server-to-server relay, and mailbox access through IMAP or POP3 (\autoref{fig:email_delivery}).
Webmail instead uses HTTPS for client-facing submission and access.
PQC on one connection cannot protect a copy captured on another connection that uses traditional TLS key establishment.
Message encryption can also protect content independently of transport, but only when implementations provide usable post-quantum encryption.
However, message-level encryption remains rare. Only 5.46\% of measured users have ever used S/MIME or PGP, and 0.06\% of emails were encrypted~\cite{stransky202227}. Thus, transport-level PQC is the primary defense against HNDL for most users, and we treat message-level PQC as a complement.

Prior email research examines classical transport security, delivery dependencies, and message encryption~\cite{durumeric2015neither,holz2015tls,holzbauer2022notsimple,li2025understanding,whitten1999johnny,muller2019johnny}.
Wickramasinghe~\etal characterize PQC support in Web-facing TLS~\cite{wickramasinghe2026mind}.
Loizou and Ghadafi compare HTTPS and the highest-priority SMTP MX hosts of 4,665 U.K.\ organizations~\cite{loizou2026measuring}.
Neither line jointly examines PQC availability across email services, negotiation by initiating peers, and usable message-level protection.

This paper targets filling this gap by measuring \textit{to what extent the email ecosystem is prepared for HNDL attacks, and where are the gaps?}
Specifically, we answer the following questions:

\begin{itemize}[leftmargin=*]
    \item \textbf{RQ1 (Server-side TLS):} How widely do email servers support PQC across submission, relay, access, and webmail, and how does support compare by provider?
    \item \textbf{RQ2 (Client-side TLS):} Do mail clients and outbound MTAs offer and negotiate PQC when receiving servers support it?
    \item \textbf{RQ3 (Message-level encryption):} Can OpenPGP and S/MIME implementations provide usable PQC message encryption independently of transport support?
\end{itemize}


To answer these questions, we combine server-side (\tlsserver) measurements with client-side (\tlsclient) tests across the email stages in \autoref{fig:email_delivery}.
For RQ1, we discover the MX, submission, access, and webmail servers associated with the Tranco Top~1M domains and test hybrid and pure ML-KEM capability through three snapshots over four weeks.
For RQ2, we capture the TLS handshakes of three widely used mail clients (Apple Mail, Gmail, and Outlook) across platforms at stages~\BC{1}, \BC{2}, and~\BC{4}.
At stage~\BC{3}, we operate a controlled MX receiving mail from eight providers under two TLS configurations, one accepting all tested groups and one accepting only PQC groups.
For RQ3, we evaluate 27 client and platform configurations, including 24 supporting OpenPGP and 17 supporting S/MIME, using a structured corpus of 19 OpenPGP keys and six S/MIME certificates.
Our main findings are as follows:

\begin{itemize}[wide, nosep, labelwidth=!, labelindent=0pt]
\item \textbf{(RQ1)} Observed PQC deployment is dominated by Google rather than broadly distributed across providers.
In the final snapshot, hybrid support reaches 33.1\% for explicit-MX relay, 40.2\% for submission, 40.2\% to 46.8\% for access, and 41.0\% for webmail within their respective stable panels.
Google accounts for approximately 87\% of hybrid-capable relay domains and 96\% to 97\% of PQC-capable submission and access domains.
Microsoft shows no PQC support at its measured endpoints, including those serving \texttt{nist.gov}, while Proton accepts none at its measured inbound MX endpoints.
Support also differs across the infrastructure associated with a domain.
Filtering gateways (e.g., Proofpoint and Mimecast) remain traditional-only even in front of PQC-capable mailbox providers, and 417 domains pair a PQC-capable primary MX with a non-PQC backup.
Mailbox-provider capability therefore does not establish consistent support across the services and relay alternatives.

\item \textbf{(RQ2)} Receiving capability does not ensure PQC use.
Only two of eight tested sending providers negotiate PQC with our receiver, while Google, Yahoo, and Mailgun accept it inbound but do not offer it in our outbound tests.
Client behavior also varies by platform.
Apple Mail on macOS and Outlook on Windows use TLS~1.2, while Outlook on macOS advertises PQC without negotiating it in the observed connections.
When our receiver restricts TLS groups to PQC but still accepts unencrypted SMTP, four of the six incompatible senders deliver in plaintext.
Software experiments identify further barriers.
Postfix advertises a hybrid group but initially supplies only a classical share, which the tested receiver selects.
Both Exim and Postfix built without ML-KEM deliver in plaintext to the tested PQC-restricted receiver under opportunistic delivery policies.
These results distinguish enabling PQC capability from ensuring its use during negotiation and delivery.

\item \textbf{(RQ3)} Message-level encryption offers only limited complement for lack of PQC-TLS. None of the 27 client configurations we examine generates a post-quantum key natively. Thus, users must create keys with external tools and import them, which raises the barrier to adoption. Whether an imported key is accepted is decided by the cryptographic backend rather than the client, so users who want to mitigate HNDL attacks must choose a client by its backend. Even a successful import does not guarantee a usable key. Only three keys, the v6 keys with a traditional primary key, complete message exchange in both GnuPG- and PGPony-based configurations. For S/MIME, although four certificate stores import ML-KEM certificates, only two encrypt to them.
\end{itemize}


\PP{Contributions}
To our knowledge, this is the first study to combine Internet-scale measurements of server-side post-quantum key-exchange support across relay, submission, access, and webmail services with controlled experiments on client-side negotiation and message-level encryption.
Specifically, we present:
\begin{itemize}[nosep,leftmargin=*]
\item A systematic measurement methodology combining server capability across email service roles, negotiation by communicating peers, and message-level interoperability, together with a three-snapshot characterization of deployment.
\item A structured corpus of 19 OpenPGP keys and six S/MIME certificates for testing algorithm and format compatibility through matched comparisons and explicit negative controls.
\item An analysis of operational barriers spanning negotiation defaults, delivery policies, and application integration, including a reported Exim defect and a GnuPG packet-encoding flaw for which correcting the output restores decryption.
\end{itemize}
\section{Background and Related Work}
\label{sec:bacground}

\subsection{Email Delivery Path}
\label{subsec:bg_path}
\autoref{fig:email_delivery} shows the service roles involved in email delivery.
We call each role a \textit{stage} and an individual transport connection a \textit{hop}.
In \textit{submission} (\BC{1}), the mail user agent (MUA) of the sender transmits a message to a mail submission agent (MSA)~\cite{rfc6409}.
In \textit{relay} (\BC{3}), the sending mail transfer agent (MTA) locates the receiving MTA through the MX records of the recipient domain and forwards the message over SMTP~\cite{rfc5321}.
A domain can list multiple MX hosts, including backup hosts used when a preferred host is unavailable.
A cloud filtering service can also receive mail before forwarding it to the mailbox provider~\cite{rao2024unfiltered,li2025understanding}.
In \textit{retrieval} (\BC{4}), the recipient accesses the mailbox through IMAP or POP3~\cite{rfc3501,rfc1939}.
Alternatively, webmail (\BC{2}) uses HTTP(S) between a browser and a webmail server, which performs submission and retrieval on behalf of the user~\cite{rfc9110}.
\autoref{tab:email_protocols_ports} in~\autoref{appx_sec:emailprotocol_tls} lists the protocols and ports.
These services can be operated by different providers and need not share a security policy~\cite{liu2021s,holzbauer2022notsimple}.
For TLS, the connection initiator is the client (\tlsclient), even when it is an outbound MTA.
The accepting endpoint is the server (\tlsserver).
This distinction motivates measuring server-side support and client-side negotiation separately.

\subsection{Confidentiality at Two Layers}
\label{subsec:bg_layers}
\PP{Transport level}
TLS protects individual connections between adjacent parties.
\textit{Explicit} TLS upgrades a plaintext connection through \texttt{STARTTLS}, whereas \textit{implicit} TLS begins when the connection is established~\cite{rfc3207,rfc2595,rfc8314}.
Protection terminates at each participating server, which processes the message before forwarding or storing it.
Because this is inherent to the email protocols rather than to TLS, we do not consider it in our HNDL threat model; message-level encryption, however, can protect the message content even at these servers.
On opportunistic SMTP relay, a sending MTA may retry without TLS after negotiation fails~\cite{durumeric2015neither,poddebniak2021starttls}.
MTA-STS and DANE provide mechanisms for requiring authenticated TLS, but do not specify a required PQC key-exchange group~\cite{rfc8461,rfc7672}.

\PP{Message level}
OpenPGP and S/MIME protect message content independently of the transport path~\cite{rfc9580,rfc8551}.
The sender encrypts the content with a fresh symmetric key and protects that key using the public key of the recipient.
In OpenPGP, the primary key binds user identities and subkeys through signatures, while encryption typically uses a dedicated subkey. Key versions define packet and fingerprint formats: v4/v6 in OpenPGP~\cite{rfc9580} and v5 in LibrePGP~\cite{librepgp}.
S/MIME uses X.509 certificates to bind users' identities to their public keys through an issuer's signature~\cite{rfc8551}.
Message-level PQC can therefore protect content even when a transport connection lacks post-quantum key establishment.
This protection requires a working sequence of key generation or import, encryption, and decryption in the participating applications.
Our evaluation concerns message content rather than the confidentiality of routing information and metadata.


\subsection{Post-Quantum Key Exchange in TLS}
\label{subsec:bg_pqtls}
In TLS~1.3, the client advertises supported key-exchange groups in \texttt{supported\_groups} and provides shares in \texttt{key\_share}~\cite{rfc8446}.
The server selects a group with a share, or requests a share for another group in \texttt{supported\_groups} through \texttt{HelloRetryRequest}, which adds a round trip.
Advertising a group therefore differs from supplying its initial share, and neither alone establishes that the group is negotiated.
We distinguish \texttt{capability} (a server accepts a PQC key exchange) from \texttt{negotiation} (a PQC group is selected in an observed handshake).
Following RFC~9794, we distinguish \textit{traditional}, \textit{hybrid}, and \textit{pure} groups~\cite{rfc9794}.
NIST standardized ML-KEM for post-quantum key establishment in 2024~\cite{nist2024fips203}.
Hybrid groups such as \texttt{X25519MLKEM768} combine a traditional elliptic-curve exchange with ML-KEM~\cite{rfc10024}.
Pure groups use ML-KEM alone and are specified separately in an Internet-Draft~\cite{ietf-tls-mlkem}.
The evaluated PQC groups use TLS~1.3.
A TLS~1.2 connection does not negotiate these groups~\cite{rfc5246}.
Our focus is key establishment and content confidentiality rather than post-quantum authentication.

\PP{HNDL threat and measurement scope}
\label{subsec:bg_hndl}
In a \textit{harvest-now, decrypt-later} (HNDL) attack, an adversary records encrypted traffic today for possible decryption using a future cryptographically relevant quantum computer~\cite{bhargavan2024formal,cisa2023quantum,sikeridis2020post}.
Protecting long-lived sensitive data can require migration before such a computer becomes available~\cite{mosca2018cybersecurity}.
Recording opportunities span client-facing and inter-provider connections, with prior work documenting passive observation of email traffic and STARTTLS interference~\cite{holz2015tls,durumeric2015neither,hoffmanandrews2014starttls}.

\begin{table}[t]
    \centering
    \caption{Scope of representative prior work.
    \CIRCLE\ : coverage, \LEFTcircle\ : partial coverage, \Circle\ : no coverage.
    Large-scale denotes a Top~1M domain population or larger, or an IPv4-wide scan.}
    \begingroup
\footnotesize
\setlength{\tabcolsep}{2.5pt}
\renewcommand{\arraystretch}{1.05}

\begin{tabular}{@{}lccccccc@{}}
\toprule
\textbf{Prior Work} &
\makecell{\textbf{About}\\\textbf{email?}} &
\makecell{\textbf{About}\\\textbf{PQC?}} &
\makecell{\textbf{Large-}\\\textbf{scale}} &
\makecell{\textbf{Multi.}\\\textbf{email}\\\textbf{hops}} &
\makecell{\textbf{MUA-}\\\textbf{side}\\\textbf{TLS}} &
\makecell{\textbf{Send.-}\\\textbf{MTA}\\\textbf{TLS}} &
\makecell{\textbf{Msg.-}\\\textbf{level}\\\textbf{enc.}} \\
\midrule
Wickramasinghe et al.~\cite{wickramasinghe2026mind} &
\Circle & \CIRCLE & \CIRCLE & \Circle & \Circle & \Circle & \Circle \\
Durumeric et al.~\cite{durumeric2015neither} &
\CIRCLE & \Circle & \CIRCLE & \CIRCLE & \Circle & \CIRCLE & \Circle \\
Holz et al.~\cite{holz2015tls} &
\CIRCLE & \Circle & \CIRCLE & \CIRCLE & \LEFTcircle & \LEFTcircle & \Circle \\
Blechschmidt and Stock~\cite{blechschmidt2023extended} &
\CIRCLE & \Circle & \CIRCLE & \CIRCLE & \Circle & \CIRCLE & \CIRCLE \\
Poddebniak et al.~\cite{poddebniak2021starttls} &
\CIRCLE & \Circle & \CIRCLE & \CIRCLE & \LEFTcircle & \Circle & \Circle \\
Loizou and Ghadafi~\cite{loizou2026measuring} &
\CIRCLE & \CIRCLE & \LEFTcircle & \Circle & \Circle & \Circle & \Circle \\
\midrule
\textbf{Our work} &
\CIRCLE & \CIRCLE & \CIRCLE & \CIRCLE & \CIRCLE & \CIRCLE & \CIRCLE \\
\bottomrule
\end{tabular}

\endgroup
    \label{tab:measurement_scope}
\end{table}

\subsection{Related Work}
\label{subsec:related}
\autoref{tab:measurement_scope} compares the scope of our study with prior work.

\PP{Post-quantum TLS}
Wickramasinghe~\etal~\cite{wickramasinghe2026mind} measure PQC support in Web-facing TLS.
Loizou and Ghadafi~\cite{loizou2026measuring} examine HTTPS and the highest-priority SMTP MX hosts of 4,665 U.K.\ organizations.
Neither study jointly covers PQC across email service roles, initiating clients, and message encryption.

\PP{Classical email security}
Prior work measures email TLS deployment and client behavior~\cite{durumeric2015neither,holz2015tls,foster2015security}.
Other studies examine STARTTLS, auto-detection, confidentiality mechanisms, and dependencies involving intermediaries and webmail~\cite{poddebniak2021starttls,tang2025multifaceted,blechschmidt2023extended,holzbauer2022notsimple,li2025understanding,li2023potential}.
These studies reveal fragmented email security but do not assess PQC deployment.

\PP{Message-level encryption}
Prior OpenPGP and S/MIME studies examine usability and implementation security~\cite{whitten1999johnny,muller2019johnny}.
We evaluate PQC key generation, import, and successful message exchange, complementing our measurements of server availability and negotiated use to identify deployment gaps across email.
\section{Study Overview}
\label{sec:method}

\PP{Overview and scope}
Our study combines two kinds of evidence that answer different questions (\autoref{fig:email_delivery}). For RQ1, we scan the submission, relay, access, and webmail endpoints associated with the Tranco Top~1M domains across three snapshots. This scan yields population-level capability rates over the endpoints we discover (\S\ref{sec:rq1_result}), so endpoints that our discovery misses are not represented. For RQ2 and RQ3, we conduct controlled experiments with mail clients, outbound MTAs, MTA software, and OpenPGP and S/MIME implementations. These experiments yield case-level observations that explain whether and why capability translates into negotiated or usable protection (\S\ref{sec:rq2_result}, \S\ref{sec:rq3_result}). The observations apply to the tested applications, platforms, providers, and configurations, and they do not estimate how often PQC is used across email traffic. Two further limits apply to both kinds of evidence. First, a PQC-capable server does not imply that PQC is used on connections to it, and our external vantage point does not observe provider-internal hops. Second, we do not consider post-quantum certificates because HNDL threatens only confidentiality: recorded traffic can be decrypted retroactively, whereas signatures cannot be forged retroactively~\cite{sikeridis2020post}.

\PP{Ethical Considerations}
\label{sec:ethics}
We follow the principles of the Menlo Report~\cite{bailey2012menlo}.
Endpoints are deduplicated across domains and probed in randomized order, with at most five short-lived connections per endpoint per snapshot, each
retried at most three times only after a timeout.
Public TLS capability probes stop after the handshake without authentication or email transmission.
We collect handshake metadata, certificates, and protocol banners.
HTTP requests identify the study through a User-Agent string linking to a project page with an opt-out option.
Controlled delivery experiments use only accounts and receiving servers under our control and do not capture unrelated traffic.
\section{RQ1: Server-Side PQC Deployment}
\label{sec:rq1_result}


\subsection{Measurement Method}
\label{subsec:rq1_method}

\PP{Domain selection and relay discovery}
We use the Tranco Top~1M snapshot of July~26, 2026~\cite{pochat2018tranco}, following the domain-based approach of prior email measurements~\cite{durumeric2015neither}.
Using \texttt{ZDNS}~v2.1.1~\cite{izhikevich2022zdns} with a 5-second timeout and up to three retries per query, we resolve MX records and the A and AAAA records supplying IPv4 and IPv6 addresses.
We exclude Null~MX domains~\cite{rfc7505}.
Domains with MX records form the explicit-MX set; otherwise, we use their A/AAAA addresses as implicit delivery targets~\cite{rfc5321}.
Domains without a delivery target are excluded.
We probe all returned addresses of each delivery target, including MX hosts at every preference tier.
An \textit{endpoint} is identified by hostname, IP address, port, and TLS mode.
\autoref{tab:mx-resolution} of Appendix~\S\ref{appx:sec:email_mx} summarizes the resolution outcomes.

\PP{Submission and access discovery}
We retrieve submission and access endpoints from Autoconfig, Autodiscover, DNS SRV records, and the Thunderbird ISPDB~\cite{wen2025automatic,autoconfig06:online}.
We do not guess hostnames for these services because a matching hostname alone does not establish an association with the domain (see~\autoref{tab:sub_access_discovery} of Appendix~\S\ref{appx:sec:sub_access_endpoing}).

\PP{Webmail discovery}
Following Li~\etal~\cite{li2023potential}, we request each domain's \texttt{mail.} and \texttt{webmail.}~\cite{ut1webmail} subdomains and follow HTTP(S) redirects to the final page.
If the redirects lead to a known provider login host (\eg, \texttt{accounts.google.com}), we attribute the domain to that provider's webmail and evaluate TLS support at the provider’s host (\eg, mail.google.com) rather than at the login host.
This rule accommodates script-rendered login pages with few indicators.
For other pages, we compute a score based on 16 webmail terms and four login-form elements (20 points), matching terms in English and seven other languages (\eg, Chinese, French).
We classify a page as webmail if it scores at least 10, or at least 7 and contains a fingerprint of webmail software (\eg, Roundcube, OWA).

\PP{TLS handshake probes}
Our OpenSSL~3.5.5 scanner~\cite{openssl35} probes each endpoint using the TLS mode listed in~\autoref{tab:email_protocols_ports} with a 7-second timeout and up to three retries per probe.
After confirming TCP reachability, we test four handshake probes in sequence: mixed, traditional-only, hybrid-only, and pure ML-KEM-only.
The mixed offer contains all groups in \autoref{tab:kex-groups} (Appendix~\S\ref{appx:sec:openssl_key}), ordered traditional, hybrid, pure.
The traditional-only probe accounts for possible incompatibility with the mixed offer.
A successful mixed or traditional-only handshake establishes TLS support, while a hybrid-only or ML-KEM-only handshake establishes PQC capability.


\PP{Provider attribution}
Following Liu~\etal~\cite{liu2021s}, we infer operators from TLS certificates (98\% of cases), protocol banners or greetings, and hostnames, in that order.
We group operators into cloud mail providers, hosting providers, and filtering services~\cite{rao2024unfiltered}.
Endpoints with certificates naming the domain itself are classified as self-hosted under this heuristic.

\PP{Support criteria and stable panels}
For each stage, we retain domains with successful TLS in all three snapshots, forming a fixed panel for temporal comparisons~\cite{wickramasinghe2026mind}.
A domain is PQC-capable at a stage if at least one discovered endpoint accepts an evaluated PQC group.
The resulting rates measure available capability, not uniform support across every alternative endpoint.
Each stage uses its own panel as the denominator, so percentages across stages do not describe the same domain population.


\subsection{PQC Adoption per Stage}
At every stage, more than half of the domains lack PQC support (\autoref{tab:stable_pqc_table_domain}), leaving them vulnerable to HNDL attacks. Their connections use traditional key exchange regardless of the client's offer, so traffic recorded today becomes decryptable once a CRQC exists. 
Among these domains, 86--94\% on the mail hops (78\% on webmail) already negotiate TLS~1.3, and the rest use only TLS~1.2.
This suggests that the lack of PQC is mostly not a matter of the TLS version, but that these servers have not enabled a hybrid group.
The domains limited to TLS~1.2 cannot negotiate PQC at all and should adopt TLS~1.3 to defend against HNDL attacks.

Since every domain supporting pure ML-KEM also supports hybrid exchange at the same stage, hybrid support alone determines the protected population. Yet one domain's support is not enough. Even a domain that supports PQC at every stage remains exposed to HNDL when it sends mail to a domain that does not, so protection requires both the sending and receiving domains to adopt PQC. Support is consistent across a hostname's addresses, and a domain's MX hosts agree in 99.0\% of cases; \S\ref{subsec:backup_mx} examines the exceptions, where a backup MX without PQC opens a downgrade path.

\begin{table}[t]
    \centering
    \caption{PQC support across services and snapshots.
    Each row uses domains with successful TLS in all three snapshots, with its denominator below the service name.
    Cells report percentages and counts.
    Hybrid and Pure indicate support observed with hybrid-only and pure ML-KEM offers.}
    \label{tab:stable_pqc_table_domain}
    \small
\setlength{\tabcolsep}{4pt}
\renewcommand{\arraystretch}{1.15}

\providecommand{\pqccell}[2]{}
\renewcommand{\pqccell}[2]{\shortstack[c]{#1\%\\(#2)}}
\providecommand{\hopcell}[2]{}
\renewcommand{\hopcell}[2]{\shortstack[c]{#1\\{\footnotesize (#2 domains)}}}

\resizebox{\linewidth}{!}{%
\begin{tabular}{c cc cc cc}
\toprule
 & \multicolumn{2}{c}{Aug 17}
 & \multicolumn{2}{c}{Aug 31}
 & \multicolumn{2}{c}{Sep 14} \\
\cmidrule(lr){2-3}
\cmidrule(lr){4-5}
\cmidrule(lr){6-7}
Hop
    & Hybrid & Pure
    & Hybrid & Pure
    & Hybrid & Pure \\
\midrule
\hopcell{Submission}{356,871}
    & \pqccell{39.8}{142,138}
    & \pqccell{38.9}{138,879}
    & \pqccell{39.9}{142,324}
    & \pqccell{38.9}{138,879}
    & \pqccell{40.2}{143,539}
    & \pqccell{38.9}{138,879} \\
\midrule

\hopcell{MX relay (explicit MX)}{508,116}
    & \pqccell{32.9}{167,295}
    & \pqccell{28.8}{146,535}
    & \pqccell{33.1}{168,194}
    & \pqccell{28.8}{146,535}
    & \pqccell{33.1}{168,319}
    & \pqccell{28.8}{146,535} \\
\addlinespace[2pt]

\hopcell{MX relay (implicit MX)}{3,552}
    & \pqccell{3.4}{119}
    & \pqccell{0.0}{0}
    & \pqccell{3.4}{120}
    & \pqccell{0.0}{0}
    & \pqccell{3.4}{119}
    & \pqccell{0.0}{0} \\
\addlinespace[2pt]
\midrule

\hopcell{Access: IMAP}{359,098}
    & \pqccell{39.8}{142,989}
    & \pqccell{38.6}{138,779}
    & \pqccell{39.8}{143,078}
    & \pqccell{38.6}{138,779}
    & \pqccell{40.2}{144,396}
    & \pqccell{38.6}{138,779} \\
\addlinespace[2pt]

\hopcell{Access: POP3}{316,683}
    & \pqccell{46.4}{146,896}
    & \pqccell{45.5}{144,053}
    & \pqccell{46.4}{146,909}
    & \pqccell{45.5}{144,053}
    & \pqccell{46.8}{148,341}
    & \pqccell{45.5}{144,053} \\
\addlinespace[2pt]
\midrule

\hopcell{Webmail}{65,904}
    & \pqccell{34.3}{22,588} & \pqccell{17.8}{11,751}
    & \pqccell{36.2}{23,834}
    & \pqccell{17.8}{11,751}
    & \pqccell{41.0}{27,017}
    & \pqccell{17.8}{11,751} \\

\bottomrule
\end{tabular}%
}
\end{table}

\subsection{PQC Adoption per Provider}
\label{subsec:rq1_pqc_provider}
\autoref{fig:provider-matrix} (Appendix~\S\ref{app:provider-domain}) groups the stable domains by provider. 
As in prior work~\cite{liu2021s}, Google, Microsoft, and Proofpoint lead MX relay.

\PP{Provider Concentration}
Google accounts for approximately 87\% of the 168,438 hybrid-capable relay domains and 96\% to 97\% of PQC-capable submission and access domains.
It is also the only named provider in \autoref{fig:provider-matrix} with observed pure ML-KEM support.
Accordingly, the pure ML-KEM support that appears consistently in the webmail rows of \autoref{tab:stable_pqc_table_domain} comes from Gmail.
In contrast, Microsoft serves approximately 103,000 to 114,000 panel domains at each measured stage, including \texttt{nist.gov}, without observed hybrid or pure ML-KEM support. Because Microsoft also operates a large share of intermediate delivery infrastructure~\cite{li2025understanding}, its lack of support likely affects more mail than its domain count alone suggests.
Proton, which markets itself as a privacy-focused provider, likewise shows no PQC support on any domains.
Support outside Google is concentrated among a few smaller providers and filtering services, with fewer than 850 self-hosted relay domains supporting hybrid exchange.

\PP{Filtering Gateway Limitations}
Most measured filtering gateways, including Proofpoint, Mimecast, and Barracuda, show no PQC support at their MX endpoints, while Cloudflare Email Security supports hybrid exchange.
A traditional-only gateway prevents PQC on the external relay connection even when the mailbox provider supports it.
For example, \texttt{zoom.us} uses Proofpoint for filtering and Google Workspace for mailboxes, as inferred in \S\ref{subsec:rq1_supporting}.
Mailbox-provider capability alone does not establish relay readiness.

\PP{Hop Heterogeneity}
PQC deployment is hop-specific even within a single provider.
Namecheap and Yahoo support hybrid exchange on all of their measured MX-relay domains but on none of their submission or access domains, whereas Fastmail shows the reverse pattern, supporting it on submission and access but not on inbound relay.
A user of these providers is thus protected on one side of the path and exposed on the other.

\PP{Provider-Driven Changes}
Over four weeks, too short to establish a trend, providers continue to adopt PQC and none withdraws support.
On relay, submission, and access, hybrid support rises by at most 0.4 percentage points, driven by single providers (Migadu on relay, Strato on submission and access), while pure ML-KEM support is unchanged.
Webmail grows faster, from 36.2\% to 41.0\% in two weeks, as independently operated hosting servers, mostly cPanel, enable hybrid exchange.
Attribution details are in \S\ref{subsec:rq1_supporting}.


\subsection{Primary and Backup MX Consistency}
\label{subsec:backup_mx}
To understand whether inconsistent support among a domain's MX hosts opens a downgrade path, we compare primary and backup MX support among the 168,319 hybrid-capable domains with explicit MX records. For 417 domains, despite the primary supporting hybrid key exchange,  a backup does not (\eg, \texttt{leboncoin.fr}, Tranco rank 1,207). Blocking access to the primary could thus force a capable sender to a classical-only backup. This connection need not violate MTA-STS or DANE if the backup satisfies the applicable authentication policy, since neither mechanism requires a PQC key-exchange group~\cite{rfc8461,rfc7672}. Conversely, for 1,222 domains, the primary lacks hybrid support while a backup supports it. These domains meet the criterion of having at least one capable endpoint, yet delivery through the preferred MX remains traditional. Assessing a domain's PQC readiness on relay therefore requires checking every MX host it lists, and domains should enable PQC on every MX host. Representative domains appear in \S\ref{subsec:rq1_supporting}.

\section{RQ2: PQC Negotiation by Clients and MTAs}
\label{sec:rq2_result}
\label{sec:method_rq2}


\subsection{Mail Client Negotiations in Practice}
\label{subsec:rq_2maliclient}

\PP{Measurement method}
To understand the behavior of PQC-TLS on the client-side in \BC{1}, \BC{2}, and \BC{4} of~\autoref{fig:email_delivery}, we tested Apple Mail, Gmail, and Outlook on the platforms in \autoref{tab:client-negotiation}, using the latest versions available as of September 2026.
These applications account for 95\% of email opens tracked by Litmus~\cite{litmus2026}. 
All mail clients are configured with our own Gmail account, whose submission and access servers accept hybrid and pure ML-KEM groups in our RQ1 scans, so that a traditional handshake can be attributed to the client rather than the server.
For each configuration, we restart the client, send and retrieve a message, and capture the TLS handshakes.
We record the TLS version, advertised \texttt{supported\_groups}, \texttt{key\_share}, and selected group to distinguish offered capability from negotiation.

\PP{Platform-Dependent Negotiation}
Even with the same application and PQC-capable server, the client platform decides whether client-facing hops resist HNDL attacks.
As shown in~\autoref{tab:client-negotiation}, Gmail on the web, Android, and iOS and Apple Mail on iOS and iPadOS offer and negotiate \texttt{X25519MLKEM768}, whereas Apple Mail on macOS and Outlook on Windows connect over TLS~1.2, which cannot offer or negotiate any PQC group.
Apple Mail thus protects the same mailbox against HNDL on an iPhone but not on a Mac.
Vendors should therefore align PQC support across all platforms of a client rather than inheriting it from each platform's TLS stack.

\PP{Intermediary-Induced Limitations}
Interestingly, Outlook on macOS offers a PQC group but does not negotiate one in any observed connection.
The \texttt{Received} headers explain why. The client connects not to the Gmail server directly but to a Microsoft relay, which supports no PQC (\S\ref{sec:rq1_result}). 
A PQC-capable client and a PQC-capable server are thus not sufficient when an intermediary the user did not configure sits between them.


\begin{table}[t]
    \centering
    \caption{PQC offers and selection in client-facing connections.
    Off. marks a PQC group in the ClientHello, and Neg. marks selection in the ServerHello.
    \CIRCLE\ means yes, \Circle\ means no, and \DashCircle\ denotes TLS~1.2 only.}
    \label{tab:client-negotiation}
    \renewcommand{\arraystretch}{1.2}
    \resizebox{0.8\columnwidth}{!}{%
    \begin{tabular}{clcccc}
        \toprule
        & & \multicolumn{2}{c}{\textbf{Submission}} & \multicolumn{2}{c}{\textbf{Access}} \\
        \cmidrule(lr){3-4} \cmidrule(lr){5-6}
        \textbf{Client} & \textbf{Platform} & \textbf{Off.} & \textbf{Neg.} & \textbf{Off.} & \textbf{Neg.} \\
        \midrule
        \multirow{4}{*}{\rotatebox[origin=c]{90}{Apple Mail}}
        & macOS v27 & \DashCircle & \DashCircle & \DashCircle & \DashCircle \\
        \cmidrule(lr){2-6}
        & iPadOS v27 & \CIRCLE & \CIRCLE & \CIRCLE & \CIRCLE \\
        \cmidrule(lr){2-6}
        & iOS v27 & \CIRCLE & \CIRCLE & \CIRCLE & \CIRCLE \\
        \midrule
        \multirow{4}{*}{\rotatebox[origin=c]{90}{Gmail}}
        & Web browser Chrome v154 & \CIRCLE & \CIRCLE & \CIRCLE & \CIRCLE \\
        \cmidrule(lr){2-6}
        & Mobile App (Android; v17)  & \CIRCLE & \CIRCLE & \CIRCLE & \CIRCLE \\
        \cmidrule(lr){2-6}
        & Mobile App (iOS; v27)  & \CIRCLE & \CIRCLE & \CIRCLE & \CIRCLE \\
        \midrule
        \multirow{2}{*}{\rotatebox[origin=c]{90}{Outlook}}
        & Desktop (Windows) v2609 & \DashCircle & \DashCircle & \DashCircle & \DashCircle \\
        \cmidrule(lr){2-6}
        & Desktop (macOS) v16.113.3 & \CIRCLE & \Circle & \CIRCLE & \Circle \\
        \bottomrule
    \end{tabular}%
    }
\end{table}
\subsection{MX Relay Negotiations in Practice}
\label{subsubsec:rq2_mxrelay}
\PP{Measurement method}
To observe the client side of the MX-relay hop (\BC{3} in~\autoref{fig:email_delivery}), we operate a controlled MX server (Postfix~3.10, OpenSSL~3.5) and send messages to it from accounts at the eight providers in \autoref{tab:outbound-mx-tls}: five with inbound PQC support that permit account registration from the USA, and three without observed inbound support.
We test two inbound-server configurations.
In the \emph{all-groups} configuration, our MX accepts every group in \autoref{tab:kex-groups}, revealing which key groups outbound MTAs offer by default.
In the \emph{PQC-only} configuration, our MX accepts only hybrid and pure ML-KEM groups, which shows whether a sender can complete a PQC handshake and how it handles failure when no traditional group is available.
For every delivery attempt, we record the groups and key shares in the \texttt{ClientHello}, the selected group, and retries.
Our test domain publishes deliberately neither MTA-STS nor DANE, which reflects the vast majority of receiving domains: fewer than 1\% of \texttt{.com}, \texttt{.net}, and \texttt{.org} domains with MX records publish TLSA records~\cite{lee2020longitudinal}, and only 0.0124\% of 1.76M scanned domains have activated MTA-STS~\cite{tatang2021first}.

\PP{Inbound--Outbound Asymmetry}
Even providers that accept PQC on their server-side do not necessarily offer it when they send, leaving the relay hop exposed to HNDL attacks in the outbound direction.
Under the all-groups configuration, only Migadu and Namecheap offer and negotiate \texttt{X25519MLKEM768} (\autoref{tab:outbound-mx-tls}).
Google, Yahoo, and Mailgun accept hybrid key exchange on their own MX hosts (\S\ref{sec:rq1_result}) but offer no PQC group in their outbound ClientHello, and Mailgun's sender still negotiates TLS~1.2.
Because the \texttt{ClientHello} of these senders contains no PQC group, the cause is not a server choice.
Therefore, even Google, the leading PQC provider on the server side, does not guarantee HNDL protection for outgoing mail.



\PP{Plaintext Fallback under PQC Enforcement}
Enforcing PQC on the receiving side to resist HNDL can backfire. Senders that cannot complete a PQC handshake may deliver the message in plaintext, exposing it to any passive eavesdropper today, not only to an adversary with a future CRQC.
Under the PQC-only configuration, the six senders without PQC split into two behaviors (\autoref{tab:outbound-mx-tls}).
Google and Proton retry the handshake and then defer the message, so delivery is delayed but never unencrypted in our experiment during our observation period (1 day).
Mailgun, however, delivers without STARTTLS after two retries, and Yahoo, Outlook, and Zoho do so within one second without any further TLS attempt.


\begin{table}[t]
    \centering
    \caption{Outbound delivery to our test MX.
    Inbound reports receiving PQC capability.
    All-groups accept traditional and PQC groups, whereas PQC-only restricts the TLS group list, not necessarily plaintext SMTP acceptance.
    \CIRCLE\ means yes, \Circle\ means no, and n/a indicates that no TLS retry was needed.}
    \label{tab:outbound-mx-tls}
    \renewcommand{\arraystretch}{1.15}
    \resizebox{\columnwidth}{!}{%
    \begin{tabular}{lcclcl}
        \toprule
        & & \multicolumn{2}{c}{\textbf{All-groups server}} & \multicolumn{2}{c}{\textbf{PQC-only server}} \\
        \cmidrule(lr){3-4} \cmidrule(lr){5-6}
        \textbf{Sender} & \textbf{Inbound} & \textbf{TLS} & \textbf{Key exchange} & \textbf{TLS retry} & \textbf{Outcome} \\
        \midrule
        Migadu & \CIRCLE & 1.3 & X25519MLKEM768 & n/a & PQC \\
        Namecheap & \CIRCLE & 1.3 & X25519MLKEM768 & n/a & PQC \\
        \midrule
        Google & \CIRCLE & 1.3 & X25519 & \CIRCLE & Defer \\
        Proton & \Circle & 1.3 & X25519 & \CIRCLE & Defer \\
        \midrule
        Mailgun & \CIRCLE & 1.2 & X25519 & \CIRCLE & Plaintext \\
        Yahoo & \CIRCLE & 1.3 & X25519 & \Circle & Plaintext \\
        Outlook & \Circle & 1.3 & P-256 & \Circle & Plaintext \\
        Zoho & \Circle & 1.3 & X25519 & \Circle & Plaintext \\
        \bottomrule
    \end{tabular}%
    }
\end{table}

\subsection{SMTP Software}
\label{subsec:smtp_software}

\PP{Measurement method}
The eight providers are large operators running proprietary or unknown MTAs.
Their behavior does not represent the population of mail servers, including the self-hosted and smaller-provider domains in \S\ref{sec:rq1_result}.
To fill this gap, we examine Exim and Postfix, which account for 55\% and 39\% of mail servers in a 2025 survey~\cite{arouna2025role}.
We test Exim~4.100.1~\cite{exim2026release41001} and Postfix~3.11.7~\cite{postfix2026release3117} on Ubuntu~26.04, each built against PQC-capable OpenSSL~3.5.5 and non-PQC OpenSSL~3.4.1.
With the PQC-capable builds, we run each implementation as both sender and receiver under two configurations: the unmodified default and a PQC-only group restriction.
With the non-PQC builds, we run each implementation as a sender toward a PQC-only receiver.
\autoref{tab:smtp-software} (\S\ref{app:smtp-software}) summarizes the results.


\PP{Default Negotiation} A default installation with a PQC-capable TLS library does not guarantee a PQC handshake. Both implementations accept hybrid exchange as receivers, and Exim also negotiates it as a sender. However, Postfix advertises \texttt{X25519MLKEM768} in \texttt{supported\_groups} while supplying a key share only for \texttt{X25519}.
A receiver that selects from the offered shares, as ours does, therefore completes the handshake with \texttt{X25519}.
Sending MTAs should therefore supply a hybrid key share, not merely advertise the group to mitigate HNDL attacks.  


\PP{Group Restrictions and Failure Handling}
A PQC-only restriction protects mail when both MTAs implement ML-KEM and apply the restriction correctly, and it causes a sender without ML-KEM to fall back to plaintext.
First, we restrict each implementation to PQC-only groups and test whether it still sends and receives mail over PQC.
Postfix behaves correctly in both directions.
Exim~4.100.1 sends in plaintext as a sender and fails to receive as a receiver because of a TLS configuration bug, which we reported and the developers have fixed.\footnote{The report link is omitted to preserve anonymity.}
With the fix, Exim also enforces the restriction in both directions.
Second, we build each implementation against OpenSSL~3.4.1, which lacks ML-KEM, and test how it behaves as a sender toward a PQC-only receiver.
Both fall back to plaintext: Exim immediately and Postfix after deferring and retrying.
These behaviors match the immediate and retry-then-plaintext fallbacks observed for providers in \S\ref{subsubsec:rq2_mxrelay}.
Because Exim and Postfix together run on 94\% of surveyed mail servers~\cite{arouna2025role}, 
plaintext fallback may affect far more mail than our provider sample shows.


\section{RQ3: Message-Level PQC Support}
\label{sec:rq3_result}


\subsection{Measurement Method}
\label{subsec:rq3_method}


\PP{Applications and test key corpus}
We evaluate 27 client--platform configurations (\autoref{tab:matrix}), selected from prior client studies~\cite{trampert2025styled,muller2019johnny} and extended with Thunderbird~(Android), each in the latest version at the time of testing. Of these, 24 support OpenPGP and 17 support S/MIME; 14 support both. The unit of analysis is the configuration rather than the client, since a client may rely on a different cryptographic backend on each platform. For example, Thunderbird uses RNP on desktop platforms and PGPony on Android.

\PP{Key corpus}
Our corpus of 19 OpenPGP keys and six S/MIME certificates (\autoref{tab:keys}) starts from A1, the minimal RFC~9980-conformant key (v6, Ed25519 primary, ML-KEM-768+X25519)~\cite{rfc9980}. 
Except for G1–G3, each key differs from at least one other corpus key in exactly one factor—key version (v4, v5), KEM (ML-KEM-1024+X448), or primary algorithm (ML-DSA, SLH-DSA, RSA-3072, legacy EdDSA)—so that a failure can be attributed to a single cause.
A1--A7 are the test vectors in Appendix~A of RFC~9980.
It also includes three GnuPG-generated LibrePGP keys (G1--G3; v2.2.2~\cite{gpgfrontend1:online}) and two combinations that RFC~9980 prohibits (N3, N4) as negative controls. The certificates O1--O6 cover ML-KEM-512/768/1024 in standard and legacy PKCS\#12 packaging; details in Appendix~\S\ref{subsec:rq3_supporting}.

\PP{Workflow tests}
For each configuration, we test whether the client generates a PQC key natively and, since users must otherwise import one, whether it imports each key, encrypts to it, and decrypts a message encrypted to it.
We count native generation as successful only if the client offers PQC key generation, an import as successful only if the keyring or certificate store lists the ML-KEM key as encryption-capable rather than reporting it as imported, and an exchange as successful only if encryption and decryption complete (\autoref{tab:matrix}).
For the disallowed keys N3 and N4, rejection is correct, and acceptance indicates non-conformance to RFC~9980.

\subsection{Native Key Generation}
None of the 27 implementations supports native PQC key generation, as their key-generation interfaces offer only RSA and ECC algorithms.
The Bat! is the closest case, delegating key generation to the GnuPG interface without implementing it.
Proton Mail, which has announced quantum-safe encryption, exposed neither a post-quantum option nor an algorithm selection in our account during the measurement period.
Consequently, users must generate PQC keys with external tools and import them manually.
This dependency reintroduces the key-management burden that two decades of usability research identify as the primary barrier to encrypted email~\cite{ruoti2019johnny,ruoti2015johnny}.
Native PQC key generation is therefore a prerequisite for usable message-level protection.

\subsection{Key Import}
\label{subsec:rq3_import}
We examine whether an externally generated PQC key can be imported.
We find that the cryptographic backend, the key version, and the primary key's signature algorithm, rather than the client itself, largely determine the import results (\autoref{tab:matrix} of Appendix~\S\ref{subsec:rq3_supporting}).

\PP{Backend Dependence}
Ten of the 24 OpenPGP configurations reject (\Circle) all 19 keys. Their backends explain this. OpenPGP.js and Bouncy Castle (C\#) have no released version that implements ML-KEM subkeys, and RNP implements them only behind a build option that is off by default and not enabled in the builds Thunderbird ships.
\footnote{OpenPGP.js~6.3.1 keeps ML-KEM in draft pull requests; RNP~0.18.1 ships with \texttt{ENABLE\_PQC} off and aborts on algorithm~35 subkeys; bc-csharp~2.70 has an open RFC~9980 feature request.}
Canary Mail bundles Bouncy Castle (Java)~1.66, which predates ML-KEM, and the macOS Keychain is closed source, so its status cannot be confirmed from documentation. 
The remaining 14 configurations use GnuPG- or PGPony-based backends and import (\LEFTcircle~or \CIRCLE) at least one key with an ML-KEM subkey.
Within each group, the results are identical. All 11 GnuPG-based configurations import the same 12 keys, and all three PGPony-based configurations import the same 15 keys.
Users who want to mitigate HNDL attacks must therefore consider a client by its backend.

\PP{Primary-Key Gating} 
ML-KEM subkey support alone does not ensure import, because the primary key's signature algorithm can block an encryption subkey. All 11 GnuPG-based configurations reject A3--A7, whose ML-DSA or SLH-DSA primary keys GnuPG cannot verify, while accepting their matched keys A1 (for A3, A5, A6) and N2 (for A4, A7), which differ only in having an Ed25519 primary key.
Since a subkey is bound to its primary key by a signature, a backend that cannot verify that signature discards the ML-KEM subkey with it. 
Unlike TLS, which negotiates key exchange independently of authentication, OpenPGP binds the encryption subkey to the primary key's signature.
The three PGPony-based configurations import A3 and A4 but fail (\LEFTcircle) to exchange messages (\S\ref{subsec:rq3_exchange}).

The PGPony-based configurations import N3, which RFC~9980 prohibits, and then fail to encrypt to it (Appendix \S\ref{subsec:rq3_supporting}).

\PP{Certificate-Store Dependence}
For S/MIME, the certificate store, not the mail client, decides whether a
certificate can be imported. Four stores accept all six certificates because
each has added ML-KEM key handling: Windows CNG (since the November 2025
update), OpenSSL~4.0.3 (ML-KEM since 3.5, CMS support since 3.6),
NSS (3.129 bundled with Thunderbird, 3.131 with Evolution), and .NET~9.0.1.
Only the first two also complete an exchange (\CIRCLE); NSS, .NET PKCS, and Horde’s OpenSSL import the certificates but cannot use them (\LEFTcircle) (see \S\ref{subsec:rq3_exchange}). The other stores reject or silently ignore the certificates because they have no ML-KEM support: gpgsm for S/MIME (GnuPG added ML-KEM to gpg, its OpenPGP program, but not to gpgsm, its S/MIME program), the iOS and macOS Keychains (Apple ships ML-KEM in CryptoKit but has not connected it to certificates), the built-in stores of three clients (closed source), and NSS~3.28.6 in Postbox (a 2017 release, before ML-KEM existed).

\subsection{Encryption and Decryption}
\label{subsec:rq3_exchange}
Successful import does not guarantee successful encryption and decryption.
For OpenPGP, most GnuPG-based configurations complete message exchange with nine of the 12 imported keys, and with all 12 after correcting the encoding flaw described below. However, PGPony-based configurations complete it with only three of the 15 keys they import.
These three, the v6 keys with a traditional primary key (A1, N2, R-v6), are the only keys that work in both GnuPG- and PGPony-based configurations, and thus the only choice for users whose correspondents may use either backend.
For S/MIME, only two of the four certificate stores that import ML-KEM certificates also encrypt to them.

\PP{GnuPG-Based Clients}
Eight of the eleven GnuPG-based configurations exchange messages with every imported key except the three v4 keys (A2, R-v4, L-v4).
For these three keys, GnuPG~2.5.24 writes twelve extra zero octets after the v4 fingerprint in the PKESK packet, so no recipient, including the sender, can decrypt the message.
Removing these octets restores decryption ($\ddagger$ in \autoref{tab:matrix}; \S\ref{appx:subsubsec:rq3_encryption_Decrpyion}).
The remaining three configurations fail in their own integration code (details in~\S\ref{appx:subsubsec:rq3_encryption_Decrpyion}).
Horde fails on every imported key because it treats a GnuPG warning about an obsolete option as an error.
Postbox fails on every v5 and v6 key because its Enigmail layer derives key IDs with the v4 rule.
Mailpile fails on G1--G3, whose v4 primary and v5 subkey fingerprints differ in length, because it loses their key records after import.

\PP{PGPony-Based Clients}
The three PGPony-based configurations exchange messages only with A1, N2, and R-v6.
The other keys they import fail because PGPony applies different logic at import and at encryption.
Import performs no version check, whereas encryption accepts only v6 keys, so every imported v4 or v5 key fails, including the prohibited N3.
Import also falls back to PGPony's own implementation, which supports ML-DSA, whereas encryption uses only bc-java, which does not, so A3 and A4 fail as well (\S\ref{appx:subsubsec:rq3_encryption_Decrpyion}).

\PP{S/MIME}
Only Outlook Classic and Mutt complete an S/MIME exchange, each with all six certificates, because only their stores both parse ML-KEM certificates (RFC~9935) and encrypt to them with \texttt{KEMRecipientInfo} (RFC~9629).
The other five configurations import the certificates and build envelopes with code that accepts only RSA and EC recipients or lacks \texttt{KEMRecipientInfo} (\S\ref{appx:subsubsec:rq3_encryption_Decrpyion}).

\section{Discussion}
\label{sec:discussion}

\subsection{Deployment Recommendations}
\label{subsec:deployment_implications}
We suggest four recommendations based on our findings.

\PP{Plaintext Fallback under PQC Enforcement.} A PQC-only receiver caused four of six non-PQC senders, and both MTAs built without ML-KEM, to deliver in plaintext (\S\ref{subsubsec:rq2_mxrelay}, \S\ref{subsec:smtp_software}). Enforcing PQC thus trades a future CRQC threat for present exposure to any passive observer. Receivers should restrict key-exchange groups only after rejecting unencrypted SMTP; senders should defer or bounce on handshake failure, as Google and Proton do, rather than retry without TLS. Because MTA-STS and DANE cannot require a key-exchange group, expressing such a policy needs an extension.

\PP{Capability–Negotiation Gap.} Three providers accept hybrid groups inbound but offer none outbound, and default Postfix advertises X25519MLKEM768 without its key share (\S\ref{subsubsec:rq2_mxrelay}, \S\ref{subsec:smtp_software}). Capability scan~\cite{loizou2026measuring} therefore overestimates protection. Providers should enable PQC on outbound MTAs and verify that the ClientHello carries the key share; readiness reports should distinguish groups a server accepts from groups observed in handshakes it initiates.

\PP{Per-Hop Coverage.} A traditional-only filtering gateway, a non-PQC backup MX (417 domains), or an unconfigured relay (Outlook on macOS) each removes PQC from a path whose endpoints support it (\S\ref{subsec:rq1_pqc_provider}, \S\ref{subsec:backup_mx}, \S\ref{subsec:rq_2maliclient}). Providers should enable PQC per service and direction, operators should verify every MX host they list, and gateway operators should treat their own support as a precondition for every domain behind them.

\PP{Message-Level Usability.} No tested client generates a PQC key natively. And the backend is crucial for determining behaviors. Clients should consider generating PQC keys natively for usability, validate imports against RFC 9980, and apply the same backend logic at import and at encryption. Until then, users should choose v6 keys with a traditional primary and an ML-KEM subkey, the only keys that worked across both tested backends.

\subsection{Limitations and Responsible Disclosure}
\label{subsec:limitations}

\PP{Infrastructure, not traffic}
Our server measurements describe infrastructure, not traffic. The Tranco sample covers ranked domains rather than email accounts, and support rates are conditional on the endpoints we discovered and on stage-specific panels with successful TLS in all snapshots. They neither share a common population across services nor establish protection along complete message paths. Service discovery and provider attribution rely on externally visible information, provider-internal connections remain unobserved, and a single vantage point over four weeks limits geographic and temporal generalization.

\PP{Case studies, not prevalence}
Our experiments are case studies, not prevalence estimates. The client, MTA, and message-level results apply to the tested applications, versions, backends, and configurations. The eight-provider sample does not estimate how often PQC is negotiated or how often senders fall back to plaintext across email traffic. Delivery outcomes are bounded by our receiver policies and observation period, and shared backends make failures across applications correlated rather than independent.

\PP{Technical barriers, not causes}
We identify technical barriers, not organizational causes. The experiments show what fails under the tested conditions, but not why providers and vendors have made their adoption decisions.
Answering this requires operator interviews or vendor surveys, which we leave for future work.

\PP{Responsible disclosure}
We reported the Exim TLS configuration bug, which the developers have fixed.
We notified the PGPony, Horde, Postbox, and Mailpile maintainers of the integration failures.
We plan to report the GnuPG PKESK encoding flaw to developers.


\section{Conclusion}
\label{sec:conclusion}
We measured PQC readiness across email services through Internet-scale server probes and controlled tests of clients, sending MTAs, and message encryption. Fewer than half of the domains at every stage support PQC, and Google accounts for approximately 87\% of hybrid-capable relay domains. 
Receiving support does not ensure outbound use, as only two of eight tested sending providers negotiate PQC despite five accepting it inbound.
Under PQC-only TLS restrictions, four of the six incompatible senders deliver in plaintext where unencrypted SMTP remains permitted. 
At the message layer, no tested client generates a PQC key natively, and key-format and integration failures prevent many imported keys from completing a message exchange. 
These findings show that migration requires more than deploying PQC algorithms.
Providers and developers must turn capability into negotiated connections and working encryption, and must ensure that handshake failures do not fall back to plaintext. \emph{Post-quantum capability is not post-quantum communication.}
\bibliographystyle{ACM-Reference-Format}
\bibliography{ref}

@String{Computing = "Computing" }

@String{Computer = "{IEEE} Computer" }

@String{Springer = "Springer-Verlag" }

@inproceedings{holzbauer2022notsimple,
author = {Florian Holzbauer and Johanna Ullrich and Martina Lindorfer and Tobias Fiebig},
title = {{Not that Simple: Email Delivery in the 21st Century}},
booktitle = {2022 USENIX Annual Technical Conference (USENIX ATC 22)},
year = {2022},
month = jul,
pages = {295--308},
address = {Carlsbad, CA},
publisher = {USENIX Association},
url = {https://www.usenix.org/conference/atc22/presentation/holzbauer}
}

@inproceedings{blechschmidt2023extended,
author = {Birk Blechschmidt and Ben Stock},
title = {{Extended Hell(o): A Comprehensive Large-Scale Study on Email Confidentiality and Integrity Mechanisms in the Wild}},
booktitle = {32nd USENIX Security Symposium (USENIX Security 23)},
year = {2023},
month = aug,
pages = {4895--4912},
address = {Anaheim, CA},
publisher = {USENIX Association},
url = {https://www.usenix.org/conference/usenixsecurity23/presentation/blechschmidt}
}

@inproceedings{poddebniak2021starttls,
author = {Damian Poddebniak and Fabian Ising and Hanno B{\"o}ck and Sebastian Schinzel},
title = {{Why TLS is Better Without STARTTLS: A Security Analysis of STARTTLS in the Email Context}},
booktitle = {30th USENIX Security Symposium (USENIX Security 21)},
year = {2021},
month = aug,
pages = {4365--4382},
publisher = {USENIX Association},
url = {https://www.usenix.org/conference/usenixsecurity21/presentation/poddebniak}
}

@article{loizou2026measuring, author = {Konstantinos Loizou and Essam Ghadafi},
title = {{Measuring Post-Quantum TLS Deployment Across UK Internet Sectors}},
journal = {arXiv preprint arXiv:2608.02147},
year = {2026},
archivePrefix = {arXiv},
eprint = {2608.02147},
primaryClass = {cs.CR},
doi = {10.48550/arXiv.2608.02147},
url = {https://arxiv.org/abs/2608.02147}
}

@article{new01,
  author  = {Peter W. Shor},
  title   = {{Polynomial-Time Algorithms for Prime Factorization and Discrete Logarithms on a Quantum Computer}},
  journal = {SIAM Journal on Computing},
  volume  = {26},
  number  = {5},
  pages   = {1484--1509},
  year    = {1997},
  doi     = {10.1137/S0097539795293172}
}

@online{hoffmanandrews2014starttls,
  author       = {Jacob Hoffman-Andrews},
  title        = {{ISPs Removing Their Customers' Email Encryption}},
  organization = {Electronic Frontier Foundation, Deeplinks Blog},
  year         = {2014},
  month        = nov,
  day          = {11},
  url          = {https://www.eff.org/deeplinks/2014/11/starttls-downgrade-attacks}
}

@misc{rfc2595,
  series       = {Request for Comments},
  number       = {2595},
  howpublished = {RFC 2595},
  publisher    = {RFC Editor},
  author       = {C. Newman},
  title        = {{Using TLS with IMAP, POP3 and ACAP}},
  year         = {1999},
  month        = jun,
  doi          = {10.17487/RFC2595},
  url          = {https://www.rfc-editor.org/info/rfc2595}
}

@ArtifactSoftware{R,
    title = {R: A Language and Environment for Statistical Computing},
    author = {{R Core Team}},
    organization = {R Foundation for Statistical Computing},
    address = {Vienna, Austria},
    year = {2019},
    url = {https://www.R-project.org/},
}

@inproceedings{izhikevich2022zdns,
  title={ZDNS: a fast DNS toolkit for internet measurement},
  author={Izhikevich, Liz and Akiwate, Gautam and Berger, Briana and Drakontaidis, Spencer and Ascheman, Anna and Pearce, Paul and Adrian, David and Durumeric, Zakir},
  booktitle={Proceedings of the 22nd ACM Internet Measurement Conference},
  pages={33--43},
  year={2022}
}

@inproceedings{wen2025automatic,
  title={Automatic Insecurity: Exploring Email Auto-configuration in the Wild.},
  author={Wen, Shushang and Zhang, Yiming and Shen, Yuxiang and Li, Bingyu and Duan, Haixin and Lin, Jingqiang},
  booktitle={NDSS},
  year={2025}
}

@article{li2023potential,
  title={The potential harm of email delivery: Investigating the HTTPS configurations of webmail services},
  author={Li, Ruixuan and Zhang, Zhenyong and Shao, Jun and Lu, Rongxing and Jia, Xiaoqi and Wei, Guiyi},
  journal={IEEE Transactions on Dependable and Secure Computing},
  volume={21},
  number={1},
  pages={125--138},
  year={2023},
  publisher={IEEE}
}

@inproceedings{bhargavan2024formal,
  title={Formal verification of the $\{$PQXDH$\}$$\{$Post-Quantum$\}$ key agreement protocol for end-to-end secure messaging},
  author={Bhargavan, Karthikeyan and Jacomme, Charlie and Kiefer, Franziskus and Schmidt, Rolfe},
  booktitle={33rd USENIX Security Symposium (USENIX Security 24)},
  pages={469--486},
  year={2024}
}

@inproceedings{rao2024unfiltered,
  title={Unfiltered: Measuring cloud-based email filtering bypasses},
  author={Rao, Sumanth and Liu, Enze and Ho, Grant and Voelker, Geoffrey M and Savage, Stefan},
  booktitle={Proceedings of the ACM Web Conference 2024},
  pages={1702--1711},
  year={2024}
}

@inproceedings{durumeric2015neither,
  title={Neither snow nor rain nor MITM... an empirical analysis of email delivery security},
  author={Durumeric, Zakir and Adrian, David and Mirian, Ariana and Kasten, James and Bursztein, Elie and Lidzborski, Nicolas and Thomas, Kurt and Eranti, Vijay and Bailey, Michael and Halderman, J Alex},
  booktitle={Proceedings of the 2015 Internet Measurement Conference},
  pages={27--39},
  year={2015}
}

@online{cisa2023quantum,
  author       = {{Cybersecurity and Infrastructure Security Agency and National Security Agency and National Institute of Standards and Technology}},
  title        = {{Quantum-Readiness: Migration to Post-Quantum Cryptography}},
  year         = {2023},
  month        = aug,
  url          = {https://www.cisa.gov/resources-tools/resources/quantum-readiness-migration-post-quantum-cryptography},
  note         = {Joint factsheet},
  urldate      = {2026-09-14}
}

@misc{rfc3207,
  series = {Request for Comments}, number = {3207}, howpublished = {RFC 3207}, publisher = {RFC Editor},
  author = {P. Hoffman},
  title  = {{SMTP Service Extension for Secure SMTP over Transport Layer Security}},
  year = {2002}, month = feb, doi = {10.17487/RFC3207}, url = {https://www.rfc-editor.org/info/rfc3207}
}

@misc{rfc7505,
  series = {Request for Comments}, number = {7505}, howpublished = {RFC 7505}, publisher = {RFC Editor},
  author = {J. Levine and M. Delany},
  title  = {{A ``Null MX'' No Service Resource Record for Domains That Accept No Mail}},
  year = {2015}, month = jun, doi = {10.17487/RFC7505}, url = {https://www.rfc-editor.org/info/rfc7505}
}

@misc{rfc8461,
  series = {Request for Comments}, number = {8461}, howpublished = {RFC 8461}, publisher = {RFC Editor},
  author = {D. Margolis and M. Risher and B. Ramakrishnan and A. Brotman and J. Jones},
  title  = {{SMTP MTA Strict Transport Security (MTA-STS)}},
  year = {2018}, month = sep, doi = {10.17487/RFC8461}, url = {https://www.rfc-editor.org/info/rfc8461}
}

@misc{rfc7672,
  series = {Request for Comments}, number = {7672}, howpublished = {RFC 7672}, publisher = {RFC Editor},
  author = {V. Dukhovni and W. Hardaker},
  title  = {{SMTP Security via Opportunistic DNS-Based Authentication of Named Entities (DANE) Transport Layer Security (TLS)}},
  year = {2015}, month = oct, doi = {10.17487/RFC7672}, url = {https://www.rfc-editor.org/info/rfc7672}
}

@misc{rfc9580,
  series = {Request for Comments}, number = {9580}, howpublished = {RFC 9580}, publisher = {RFC Editor},
  author = {P. Wouters and D. Huigens and J. Winter and Y. Niibe},
  title  = {{OpenPGP}},
  year = {2024}, month = jul, doi = {10.17487/RFC9580}, url = {https://www.rfc-editor.org/info/rfc9580}
}

@misc{rfc8551,
  series = {Request for Comments}, number = {8551}, howpublished = {RFC 8551}, publisher = {RFC Editor},
  author = {J. Schaad and B. Ramsdell and S. Turner},
  title  = {{Secure/Multipurpose Internet Mail Extensions (S/MIME) Version 4.0 Message Specification}},
  year = {2019}, month = apr, doi = {10.17487/RFC8551}, url = {https://www.rfc-editor.org/info/rfc8551}
}

@misc{rfc8446,
  series = {Request for Comments}, number = {8446}, howpublished = {RFC 8446}, publisher = {RFC Editor},
  author = {E. Rescorla},
  title  = {{The Transport Layer Security (TLS) Protocol Version 1.3}},
  year = {2018}, month = aug, doi = {10.17487/RFC8446}, url = {https://www.rfc-editor.org/info/rfc8446}
}

@misc{rfc5246,
  series = {Request for Comments}, number = {5246}, howpublished = {RFC 5246}, publisher = {RFC Editor},
  author = {T. Dierks and E. Rescorla},
  title  = {{The Transport Layer Security (TLS) Protocol Version 1.2}},
  year = {2008}, month = aug, doi = {10.17487/RFC5246}, url = {https://www.rfc-editor.org/info/rfc5246}
}

@misc{ietf-tls-mlkem,
  author       = {D. Connolly},
  title        = {{ML-KEM Post-Quantum Key Agreement for TLS 1.3}},
  howpublished = {Internet-Draft, draft-ietf-tls-mlkem, IETF},
  note         = {Work in progress},
  year         = {2025},
  url          = {https://datatracker.ietf.org/doc/draft-ietf-tls-mlkem/}
}

@inproceedings{whitten1999johnny,
  author    = {Alma Whitten and J. D. Tygar},
  title     = {{Why Johnny Can't Encrypt: A Usability Evaluation of PGP 5.0}},
  booktitle = {Proceedings of the 8th USENIX Security Symposium},
  year      = {1999}
}

@article{mosca2018cybersecurity,
  author  = {Michele Mosca},
  title   = {{Cybersecurity in an Era with Quantum Computers: Will We Be Ready?}},
  journal = {IEEE Security \& Privacy},
  volume  = {16},
  number  = {5},
  pages   = {38--41},
  year    = {2018},
  doi     = {10.1109/MSP.2018.3761723}
}

@misc{rfc8314,
  series       = {Request for Comments},
  number       = {8314},
  howpublished = {RFC 8314},
  publisher    = {RFC Editor},
  author       = {K. Moore and C. Newman},
  title        = {{Cleartext Considered Obsolete: Use of Transport Layer Security (TLS) for Email Submission and Access}},
  year         = {2018},
  month        = jan,
  doi          = {10.17487/RFC8314},
  url          = {https://www.rfc-editor.org/info/rfc8314}
}

@misc{rfc5321,
  series       = {Request for Comments},
  number       = {5321},
  howpublished = {RFC 5321},
  publisher    = {RFC Editor},
  author       = {J. Klensin},
  title        = {{Simple Mail Transfer Protocol}},
  year         = {2008},
  month        = oct,
  doi          = {10.17487/RFC5321},
  url          = {https://www.rfc-editor.org/info/rfc5321}
}

@misc{rfc6409,
  series       = {Request for Comments},
  number       = {6409},
  howpublished = {RFC 6409},
  publisher    = {RFC Editor},
  author       = {R. Gellens and J. Klensin},
  title        = {{Message Submission for Mail}},
  year         = {2011},
  month        = nov,
  doi          = {10.17487/RFC6409},
  url          = {https://www.rfc-editor.org/info/rfc6409}
}

@misc{rfc3501,
  series       = {Request for Comments},
  number       = {3501},
  howpublished = {RFC 3501},
  publisher    = {RFC Editor},
  author       = {M. Crispin},
  title        = {{Internet Message Access Protocol - Version 4rev1}},
  year         = {2003},
  month        = mar,
  doi          = {10.17487/RFC3501},
  url          = {https://www.rfc-editor.org/info/rfc3501}
}

@misc{rfc1939,
  series       = {Request for Comments},
  number       = {1939},
  howpublished = {RFC 1939},
  publisher    = {RFC Editor},
  author       = {J. Myers and M. Rose},
  title        = {{Post Office Protocol - Version 3}},
  year         = {1996},
  month        = may,
  doi          = {10.17487/RFC1939},
  url          = {https://www.rfc-editor.org/info/rfc1939}
}

@misc{rfc9110,
  series       = {Request for Comments},
  number       = {9110},
  howpublished = {RFC 9110},
  publisher    = {RFC Editor},
  author       = {R. Fielding and M. Nottingham and J. Reschke},
  title        = {{HTTP Semantics}},
  year         = {2022},
  month        = jun,
  doi          = {10.17487/RFC9110},
  url          = {https://www.rfc-editor.org/info/rfc9110}
}

@inproceedings{liu2021s,
  title={Who's got your mail? characterizing mail service provider usage},
  author={Liu, Enze and Akiwate, Gautam and Jonker, Mattijs and Mirian, Ariana and Savage, Stefan and Voelker, Geoffrey M},
  booktitle={Proceedings of the 21st ACM Internet Measurement Conference},
  pages={122--136},
  year={2021}
}

@inproceedings{wickramasinghe2026mind,
  title={Mind the gap: Policy vs reality in post-quantum TLS deployment},
  author={Wickramasinghe, Nimesha and Li, Frank and Jha, Sanjay and Shaghaghi, Arash},
  booktitle={Proceedings of the 2026 ACM Internet Measurement Conference},
  pages={256--275},
  year={2026}
}

@online{statista2025emails,
  author  = {{Statista Research Department}},
  title   = {Number of sent and received e-mails per day worldwide from 2018 to 2028 (in billions)},
  year    = {2025},
  month   = {9},
  url     = {https://www.statista.com/statistics/456500/daily-number-of-e-mails-worldwide/},
  urldate = {2026-08-25},
  organization = {Statista}
}

@misc{nist2024fips203,
author = {{National Institute of Standards and Technology}},
title = {{Module-Lattice-Based Key-Encapsulation Mechanism Standard}},
institution = {National Institute of Standards and Technology},
type = {Federal Information Processing Standards Publication},
number = {203},
year = {2024},
month = aug,
doi = {10.6028/NIST.FIPS.203},
url = {https://doi.org/10.6028/NIST.FIPS.203}
}

@misc{rfc10024,
series = {Request for Comments},
number = {10024},
howpublished = {RFC 10024},
publisher = {RFC Editor},
author = {K. Kwiatkowski and P. Kampanakis and B. E. Westerbaan and D. Stebila},
title = {{Post-Quantum Traditional (PQ/T) Hybrid Key Agreement Mechanisms for TLS 1.3}},
year = {2026},
month = aug,
pagetotal = {9},
doi = {10.17487/RFC10024},
url = {https://www.rfc-editor.org/info/rfc10024/}
}

@misc{rfc9958,
  author       = {Banerjee, Aritra and Reddy K., Tirumaleswar and
                  Schoinianakis, Dimitrios and Hollebeek, Timothy and
                  Ounsworth, Mike},
  title        = {{Post-Quantum Cryptography for Engineers}},
  series       = {Request for Comments},
  number       = {9958},
  howpublished = {RFC 9958},
  publisher    = {RFC Editor},
  year         = {2026},
  month        = jun,
  doi          = {10.17487/RFC9958},
  url          = {https://www.rfc-editor.org/info/rfc9958}
}

@online{autoconfig06:online,
  author       = {{Ben Bucksch}},
  title        = {Mail Autoconfig: draft-ietf-mailmaint-autoconfig-06},
  year         = {2026},
  month        = jun,
  url          = {https://datatracker.ietf.org/doc/draft-ietf-mailmaint-autoconfig/06/},
  note         = {autoconfig-06},
  urldate      = {2026-08-25}
}

@inproceedings{tang2025multifaceted,
  title={A Multifaceted Study on the Use of TLS and Auto-detect in Email Ecosystems.},
  author={Tang, Ka Fun and Tu, Che Wei and Mak, Sui Ling Angela and Chau, Sze Yiu},
  booktitle={NDSS},
  year={2025}
}

@article{holz2015tls,
  title={TLS in the wild: An Internet-wide analysis of TLS-based protocols for electronic communication},
  author={Holz, Ralph and Amann, Johanna and Mehani, Olivier and Wachs, Matthias and Kaafar, Mohamed Ali},
  journal={arXiv preprint arXiv:1511.00341},
  year={2015}
}

@inproceedings{foster2015security,
  title={Security by any other name: On the effectiveness of provider based email security},
  author={Foster, Ian D and Larson, Jon and Masich, Max and Snoeren, Alex C and Savage, Stefan and Levchenko, Kirill},
  booktitle={Proceedings of the 22nd ACM SIGSAC conference on computer and communications security},
  pages={450--464},
  year={2015}
}

@inproceedings{li2025understanding,
  title={Understanding and characterizing intermediate paths of email delivery: The hidden dependencies},
  author={Li, Ruixuan and Lu, Chaoyi and Liu, Baojun and Lin, Yanzhong and Duan, Haixin and Pan, Qingfeng and Shao, Jun},
  booktitle={Proceedings of the 2025 ACM Internet Measurement Conference},
  pages={578--591},
  year={2025}
}

@article{pochat2018tranco,
  title={Tranco: A research-oriented top sites ranking hardened against manipulation},
  author={Pochat, Victor Le and Van Goethem, Tom and Tajalizadehkhoob, Samaneh and Joosen, Wouter and others},
  journal={arXiv preprint arXiv:1806.01156},
  year={2018}
}

@misc{openssl35,
  author       = {{OpenSSL Library}},
  title        = {{OpenSSL} 3.5 Series Release Notes},
  howpublished = {\url{https://openssl-library.org/news/openssl-3.5-notes/}},
  note         = {Accessed: September 16, 2026}
}

@inproceedings{trampert2025styled,
  title={Styled to Steal: The Overlooked Attack Surface in Email Clients},
  author={Trampert, Leon and Weber, Daniel and Rossow, Christian and Schwarz, Michael},
  booktitle={Proceedings of the 2025 ACM SIGSAC Conference on Computer and Communications Security},
  pages={2069--2083},
  year={2025}
}

@inproceedings{muller2019johnny,
  title={$\{$“Johnny$\}$, you are $\{$fired!”$\}$--Spoofing $\{$OpenPGP$\}$ and $\{$S/MIME$\}$ Signatures in Emails},
  author={M{\"u}ller, Jens and Brinkmann, Marcus and Poddebniak, Damian and B{\"o}ck, Hanno and Schinzel, Sebastian and Somorovsky, Juraj and Schwenk, J{\"o}rg},
  booktitle={28th USENIX Security Symposium (USENIX Security 19)},
  pages={1011--1028},
  year={2019}
}

@misc{rfc9794,
  series       = {Request for Comments},
  number       = 9794,
  howpublished = {RFC 9794},
  publisher    = {RFC Editor},
  doi          = {10.17487/RFC9794},
  url          = {https://www.rfc-editor.org/info/rfc9794},
  author       = {F. Driscoll and M. Parsons and B. Hale},
  title        = {{Terminology for Post-Quantum Traditional Hybrid Schemes}},
  year         = 2025,
  month        = jun
  }

@misc{ut1webmail,
  author       = {{Universit{\'e} Toulouse 1 Capitole}},
  title        = {{UT1} Web Categorization Blacklists: Webmail Category},
  howpublished = {\url{https://github.com/olbat/ut1-blacklists}},
  year         = {2026},
  note         = {Accessed: September 26, 2026}
}

@misc{litmus2026,
  author       = {{Litmus}},
  title        = {Email Client Market Share: Trends \& Statistics 2026},
  howpublished = {\url{https://www.litmus.com/email-client-market-share/}},
  year         = {2026},
  note         = {July 2026 report; data current as of August 1, 2026. Accessed: September 27, 2026}
}

@article{bailey2012menlo,
  title={The menlo report},
  author={Bailey, Michael and Dittrich, David and Kenneally, Erin and Maughan, Doug},
  journal={IEEE Security \& Privacy},
  volume={10},
  number={2},
  pages={71--75},
  year={2012},
  publisher={IEEE}
}

@inproceedings{arouna2025role,
  title={On the Role of Forward-Confirmed reverse DNS in E-mail Authentication},
  author={Arouna, Alfred and Fontein, Rick and Meijerink, Bernd and Livadariu, Ioana and Jonker, Mattijs},
  booktitle={2025 9th Network Traffic Measurement and Analysis Conference (TMA)},
  pages={1--11},
  year={2025},
  organization={IEEE}
}

@misc{exim2026release41001,
  author       = {{toolsmith}},
  title        = {{[exim-dev] exim-4.100.1 in exim/exim released}},
  year         = {2026},
  month        = sep,
  howpublished = {Exim-dev mailing list},
  url          = {https://www.mail-archive.com/exim-dev@lists.exim.org/msg01744.html},
  note         = {Accessed: October 4, 2026}
}

@misc{postfix2026release3117,
  author = {Wietse Venema},
  title  = {{Postfix} stable release 3.11.7 and legacy releases
            3.10.14, 3.9.15, 3.8.21, 3.7.23, 3.6.21, 3.5.28},
  year   = {2026},
  month  = sep,
  url    = {https://www.postfix.org/announcements/postfix-3.11.7.html},
  note   = {Accessed: October 4, 2026}
}

@article{sikeridis2020post,
  title={Post-quantum authentication in TLS 1.3: A performance study},
  author={Sikeridis, Dimitrios and Kampanakis, Panos and Devetsikiotis, Michael},
  journal={Cryptology ePrint Archive},
  year={2020}
}

@inproceedings{tatang2021first,
  title={A first large-scale analysis on usage of mta-sts},
  author={Tatang, Dennis and Flume, Robin and Holz, Thorsten},
  booktitle={International Conference on Detection of Intrusions and Malware, and Vulnerability Assessment},
  pages={361--370},
  year={2021},
  organization={Springer}
}

@inproceedings{lee2020longitudinal,
  title={A longitudinal and comprehensive study of the $\{$dane$\}$ ecosystem in email},
  author={Lee, Hyeonmin and Gireesh, Aniketh and van Rijswijk-Deij, Roland and Chung, Taejoong and others},
  booktitle={29th $\{$USENIX$\}$ Security Symposium ($\{$USENIX$\}$ Security 20)},
  year={2020}
}

@article{ruoti2015johnny,
  title={Why Johnny still, still can't encrypt: Evaluating the usability of a modern PGP client},
  author={Ruoti, Scott and Andersen, Jeff and Zappala, Daniel and Seamons, Kent},
  journal={arXiv preprint arXiv:1510.08555},
  year={2015}
}

@article{ruoti2019johnny,
  title={Johnny's journey toward usable secure email},
  author={Ruoti, Scott and Seamons, Kent},
  journal={IEEE Security \& Privacy},
  volume={17},
  number={6},
  pages={72--76},
  year={2019},
  publisher={IEEE}
}

@techreport{rfc9980,
  author    = {Kousidis, Stavros and Roth, Johannes and Strenzke, Falko and Wussler, Aron},
  title     = {Post-Quantum Cryptography in {OpenPGP}},
  howpublished = {RFC 9980},
  type      = {RFC},
  number    = {9980},
  institution = {Internet Engineering Task Force},
  publisher = {RFC Editor},
  month     = jun,
  year      = {2026},
  doi       = {10.17487/RFC9980},
  url       = {https://www.rfc-editor.org/info/rfc9980},
  note      = {Proposed Standard}
}

@techreport{librepgp,
  author      = {Werner Koch and Ronald Henry Tse},
  title       = {{LibrePGP Message Format}},
  type        = {Internet-Draft},
  number      = {draft-koch-librepgp-06},
  institution = {Internet Engineering Task Force},
  year        = {2026},
  month       = sep,
  note        = {Work in Progress},
  url         = {https://datatracker.ietf.org/doc/draft-koch-librepgp/06/}
}

@inproceedings{stransky202227,
  title={27 years and 81 million opportunities later: Investigating the use of email encryption for an entire university},
  author={Stransky, Christian and Wiese, Oliver and Roth, Volker and Acar, Yasemin and Fahl, Sascha},
  booktitle={2022 IEEE Symposium on Security and Privacy (SP)},
  pages={860--875},
  year={2022},
  organization={IEEE}
}

@misc{gpgfrontend1:online,
  author       = {Saturneric},
  title        = {{GpgFrontend}: A Cross-Platform {OpenPGP} Tool with
                  {GnuPG} and {rPGP} Backends},
  year         = {2026},
  url          = {https://github.com/saturneric/GpgFrontend},
  note         = {Accessed on October 26, 2026}
}

\appendix

\section{Email Protocols and TLS Modes}
\label{appx_sec:emailprotocol_tls}
Table~\ref{tab:email_protocols_ports} lists the protocols, ports, and TLS modes of the email services described in \S\ref{subsec:bg_path}.

\begin{table}[h]
    \centering
    \caption{Protocols, ports, and TLS modes of email services.}
    \resizebox{0.8\columnwidth}{!}{%
\begin{tabular}{cccc}
\toprule
\textbf{Protocol} & \textbf{Port} & \textbf{Service} & \textbf{TLS Mode} \\
\midrule
\multirow{3}{*}{SMTP} & 25~\cite{rfc5321}  & Relay and submission        & Explicit  \\
                      & 587~\cite{rfc6409} & \multirow{2}{*}{Submission} & Explicit  \\
                      & 465~\cite{rfc8314} &                             & Implicit \\
\midrule
\multirow{2}{*}{IMAP} & 143~\cite{rfc3501} & \multirow{2}{*}{Access}     & Explicit  \\
                      & 993~\cite{rfc8314} &                             & Implicit \\
\midrule
\multirow{2}{*}{POP3} & 110~\cite{rfc1939} & \multirow{2}{*}{Access}     & Explicit  \\
                      & 995~\cite{rfc8314} &                             & Implicit \\
\midrule
\multirow{2}{*}{HTTP} & 80~\cite{rfc9110}  & \multirow{2}{*}{Webmail}    & None \\
                      & 443~\cite{rfc9110} &                             & Implicit \\
\bottomrule
\end{tabular}%
}
    \label{tab:email_protocols_ports}
\end{table}

\section{Email Infrastructure Discovery}
\label{appx:sec:email_mx}
\label{appx:sec:sub_access_endpoing}

Table~\ref{tab:mx-resolution} shows how the Tranco Top~1M domains resolve to receiving infrastructure via MX records, while Table~\ref{tab:sub_access_discovery} shows how the submission and access endpoints for those domains are discovered following the methodology of~\cite{wen2025automatic}.  540,650 of the 666,181 domains with a valid MX endpoint complete a TLS handshake on the relay hop, as do 366,955 of 385,135 domains on submission and 374,680 of 388,161 on access (IMAP and POP3). For webmail, 185,640 domains return a page on their \texttt{mail.} or \texttt{webmail.} host, of which 71,102 remain after verification and 69,150 complete a TLS handshake.

\begin{table}[h]
    \centering
    \caption{DNS resolution outcomes for the Tranco Top~1M domains.
    Percentages use all one million domains, and indented rows subdivide the preceding category.}
    \label{tab:mx-resolution}
    \scalebox{0.85}{%
    \begin{tabular}{lr}
        \toprule
        \textbf{Type} & \textbf{Count (\%)} \\
        \midrule
        Explicit MX & 666,226 \phantom{0}(66.6\%) \\
        $\hookrightarrow$ \textbf{Valid MX} & \textbf{666,181} \phantom{0}\textbf{(66.6\%)} \\
        $\hookrightarrow$ Malformed MX & 45\phantom{0}\phantom{0} (0.0\%) \\
        No MX & 304,515 \phantom{0}(30.5\%) \\
        $\hookrightarrow$ \textbf{Implicit MX} & \textbf{250,545 \phantom{0}(25.1\%)} \\
        $\hookrightarrow$ No A/AAAA record & 53,970 \phantom{0}\phantom{0}(5.4\%) \\
        Null MX & 4,327 \phantom{0}\phantom{0}(0.4\%) \\
        NXDOMAIN & 10,674 \phantom{0}(1.1\%) \\
        DNS error & 14,258 \phantom{0}(1.4\%) \\
        \midrule
        Total & 1,000,000 (100.0\%) \\
        \bottomrule
    \end{tabular}
    }
\end{table}


\begin{table}[h]
    \centering
    \caption{Endpoint discovery for the client-facing hops (Submissions and Access).}
    \label{tab:sub_access_discovery}
    \resizebox{0.85\columnwidth}{!}{%
    \begin{tabular}{lrrr}
        \toprule
        \multirow{2}{*}{\textbf{Source}} & \multirow{2}{*}{\textbf{Submission}} & \multicolumn{2}{c}{\textbf{Access}} \\
        \cmidrule(lr){3-4}
         & & \textbf{IMAP} & \textbf{POP3} \\
        \midrule
        Autoconfig   &  30,366 \phantom{0}(7.9\%) &  27,866 \phantom{0}(7.2\%) &  22,410 \phantom{0}(6.8\%) \\
        Autodiscover &  57,537 (14.9\%) &  56,770 (14.7\%) &   8,016 \phantom{0}(2.4\%) \\
        DNS SRV      &   7,354 \phantom{0}(1.9\%) &  11,464 \phantom{0}(3.0\%) &   7,516 \phantom{0}(2.3\%) \\
        ISPDB        & 289,878 (75.3\%) & 289,127 (75.1\%) & 293,157 (88.5\%) \\
        \midrule
        \textbf{Domains} & \textbf{385,135 (100\%)} & \textbf{385,227 (100\%)} & \textbf{331,099 (100\%)} \\
        \bottomrule
    \end{tabular}%
    }
\end{table}

\section{Key Exchange Groups in OpenSSL v3.5.}
\label{appx:sec:openssl_key}

\begin{table}[th]
\centering
\caption{Key exchange groups by category in OpenSSL v3.5.}
\label{tab:kex-groups}

\begin{tabularx}{\linewidth}{llX}
\toprule
\textbf{Category} & \textbf{\#} & \multicolumn{1}{c}{\textbf{Groups}} \\
\midrule
Traditional & 13 & x25519, secp256r1, secp384r1, secp521r1,  brainpoolP256r1tls13, brainpoolP384r1tls13,   brainpoolP512r1tls13, ffdhe2048, ffdhe3072, ffdhe4096, ffdhe6144, ffdhe8192, x448 \\
\midrule
Hybrid & 3 & X25519MLKEM768, SecP256r1MLKEM768,  SecP384r1MLKEM1024 \\
\midrule
Pure & 3 & MLKEM512, MLKEM768, MLKEM1024 \\

\bottomrule
\end{tabularx}
\end{table}

\autoref{tab:kex-groups} lists the 19 key exchange groups supported by OpenSSL v3.5, grouped into three categories: 13 traditional, 3 hybrid, and 3 pure post-quantum groups.

\section{PQC Support by Provider and Service}
\label{app:provider-domain}
\begin{figure}[ht]
    \centering
    \includegraphics[width=0.85\columnwidth]{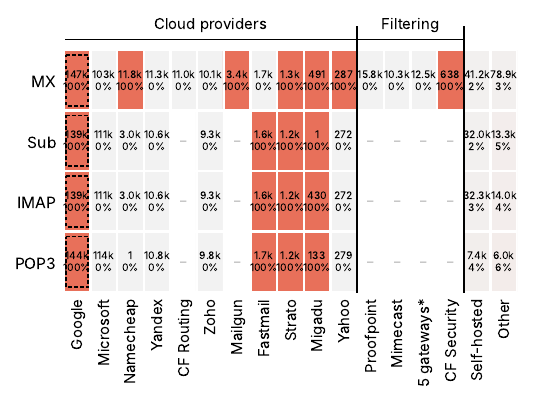}
    \caption{PQC support by provider and service. Numbers give domain counts, and colors show the fraction with at least one hybrid-capable endpoint of that provider. Dashed markings indicate pure ML-KEM support. 5 gateways$^*$: N-able, Barracuda, TrendMicro, Hornetsecurity, and Cisco.}
    \label{fig:provider-matrix}
\end{figure}

Figure~\ref{fig:provider-matrix} compares hybrid PQC support across providers and service stages, supporting the provider-level findings in \S\ref{subsec:rq1_pqc_provider}.

\section{PQC Behavior of SMTP Software}
\label{app:smtp-software}
Table~\ref{tab:smtp-software} summarizes how Exim and Postfix negotiate hybrid ML-KEM groups as receivers and senders under default and PQC-only TLS configurations, and how a sender without ML-KEM support behaves toward a PQC-only receiver.

\begin{table}[t]
    \caption{PQC behavior of Exim and Postfix as receivers (Recv.) and senders (Send.).
    Default uses the unmodified configuration, and PQC-only restricts TLS groups.
    \CIRCLE\ denotes hybrid negotiation, \LEFTcircle\ an advertised hybrid group without its initial share, and \Circle\ no support.
    The final column reports delivery to the test receiver without sender-side ML-KEM support.}
    \label{tab:smtp-software}
    \centering
    \fontsize{7.5pt}{9pt}\selectfont
    \setlength{\tabcolsep}{0pt}
    \begin{tabular*}{\columnwidth}{@{\extracolsep{\fill}} l c c c c l @{}}
        \toprule
        & \multicolumn{4}{c}{\textbf{OpenSSL 3.5.5}} & \textbf{OpenSSL 3.4.1} \\
        \cmidrule{2-5}\cmidrule{6-6}
        & \multicolumn{2}{c}{Default} & \multicolumn{2}{c}{PQC-only} & Send.\ to a \\
        \textbf{Software} & Recv. & Send. & Recv. & Send. & PQC-only receiver \\
        \midrule
        Exim 4.100.1 & \CIRCLE & \CIRCLE & \CIRCLE\rlap{$^{\dagger}$} & \CIRCLE\rlap{$^{\dagger}$} & Plaintext \\
        Postfix 3.11.7 & \CIRCLE & \LEFTcircle\rlap{$^{\ddagger}$} & \CIRCLE & \CIRCLE & Deferred, then plaintext\rlap{$^{\S}$} \\
        \bottomrule
    \end{tabular*}

    \vspace{3pt}
    \parbox{\columnwidth}{\fontsize{7pt}{8.4pt}\selectfont
    $^{\dagger}$\,Results after the reported fix.
    The original release sends plaintext or fails to receive under the restriction.\quad
    $^{\ddagger}$\,Advertises X25519MLKEM768 but initially supplies only an X25519 share.\quad
    $^{\S}$\,Plaintext on a retry.}
\end{table}

\section{Supporting Measurement Details}
\label{sec:rq_supporting}
\begin{table}[tbp]
\centering
\caption{The 25 keys and certificates used in the measurements, arranged as a primary~$\times$~key-version~$\times$~KEM grid. 
A1 is the baseline; except for the G keys, every OpenPGP key differs from at least one other key in the grid in exactly one factor (\eg, A4 and A7 differ from N2 only in the primary algorithm).
A1--A7 are the RFC~9980 Appendix~A test vectors; we generated the N, R, and L keys, and GnuPG the G keys. \emph{ID} is the OpenPGP algorithm identifier of the encryption subkey. A dash marks a combination not in the corpus.}
\label{tab:keys}
\scriptsize
\setlength{\tabcolsep}{3pt}
\begin{tabular}{@{}c l l l l@{}}
\toprule
 & & & ML-KEM-768 & ML-KEM-1024 \\
 & Primary algorithm & Version & (+X25519, ID 35) & (+X448, ID 36) \\
\midrule
\multirow{5}{*}{\rotatebox[origin=c]{90}{Traditional}}
 & Ed25519 & v6 & A1 (baseline) & N2 \\
 & Ed25519 & v5$^{a}$ & N1 & N3$^{b}$ \\
 & Ed25519 & v4 & A2 & N4$^{b}$ \\
 & RSA-3072 & v6/v5$^{a}$/v4 & R-v6/R-v5/R-v4 & -- \\
 & EdDSALegacy & v5$^{a}$/v4 & L-v5/L-v4 & --$^{c}$ \\
\midrule
\multirow{5}{*}{\rotatebox[origin=c]{90}{Post-quantum}}
 & ML-DSA-65+Ed25519 & v6 & A3 & -- \\
 & ML-DSA-87+Ed448 & v6 & -- & A4 \\
 & SLH-DSA-SHAKE-128s & v6 & A5 & -- \\
 & SLH-DSA-SHAKE-128f & v6 & A6 & -- \\
 & SLH-DSA-SHAKE-256s & v6 & -- & A7 \\
\midrule
\multicolumn{2}{@{}l}{\textit{LibrePGP}$^{d}$ EdDSA-legacy} & v4+v5 & G1, G2 & G3 \\
\midrule
\multicolumn{5}{@{}l}{\textit{S/MIME} (X.509): O1--O6 = ML-KEM-512/768/1024 $\times$ std./legacy PKCS\#12} \\
\bottomrule
\end{tabular}
\par\smallskip
\begin{minipage}{\columnwidth}
\scriptsize
$^{a}$~v5 is not defined by RFC~9580: the format GnuPG uses in place of v6.
$^{b}$~Prohibited by RFC~9980 (\S3.5): negative control, so rejection is expected.
$^{c}$~RFC~9580 does not permit EdDSALegacy in v6 keys: a new v6 key uses Ed25519 (A1).
$^{d}$~Generated by GnuPG: v4 primary with a v5 subkey, LibrePGP algorithm ID~8; the hybrid's traditional component is X25519 (G1), P-256 (G2), and P-384 (G3).  
\end{minipage}
\end{table}
%
\begin{table*}[t]
\centering
\caption{Per-client support for post-quantum OpenPGP and S/MIME keys. \CIRCLE~= key imported and messages exchanged end to end; \LEFTcircle~= key imported but sending or receiving failed; \Circle~= key import failed; $^{\ddagger}$~= exchange succeeds only after correcting the PKESK encoding of GnuPG~2.5.24 (RQ3); n/a~= track not applicable to that client.}
\label{tab:matrix}
\scriptsize
\setlength{\tabcolsep}{3pt}
\renewcommand{\arraystretch}{1.05}
\resizebox{\textwidth}{!}{%
\begin{tabular}{@{}ll l *{16}{c} *{3}{c} *{6}{c}@{}}
\toprule
& & & \multicolumn{16}{c}{OpenPGP -- RFC 9980 (PQC)} & \multicolumn{3}{c}{GnuPG} & \multicolumn{6}{c}{S/MIME (X.509)} \\
\cmidrule(lr){4-19}\cmidrule(lr){20-22}\cmidrule(l){23-28}
Platform & Client & OpenPGP Backend / S/MIME Backend & A1 & A2 & A3 & A4 & A5 & A6 & A7 & N1 & N2 & N3 & N4 & R-v4 & R-v5 & R-v6 & L-v4 & L-v5 & G1 & G2 & G3 & O1 & O2 & O3 & O4 & O5 & O6 \\
\midrule
Windows & eM Client & BouncyCastle Csharp 2.70 / .NET Pkcs 9.0.1 & \Circle & \Circle & \Circle & \Circle & \Circle & \Circle & \Circle & \Circle & \Circle & \Circle & \Circle & \Circle & \Circle & \Circle & \Circle & \Circle & \Circle & \Circle & \Circle & \LEFTcircle & \LEFTcircle & \LEFTcircle & \LEFTcircle & \LEFTcircle & \LEFTcircle \\
 & The Bat! & GnuPG 2.5.24 / built-in & \CIRCLE & \CIRCLE\rlap{$^{\ddagger}$} & \Circle & \Circle & \Circle & \Circle & \Circle & \CIRCLE & \CIRCLE & \Circle & \Circle & \CIRCLE\rlap{$^{\ddagger}$} & \CIRCLE & \CIRCLE & \CIRCLE\rlap{$^{\ddagger}$} & \CIRCLE & \CIRCLE & \CIRCLE & \CIRCLE & \Circle & \Circle & \Circle & \Circle & \Circle & \Circle \\
 & Outlook classic & GnuPG 2.5.24 / Windows CNG & \CIRCLE & \CIRCLE\rlap{$^{\ddagger}$} & \Circle & \Circle & \Circle & \Circle & \Circle & \CIRCLE & \CIRCLE & \Circle & \Circle & \CIRCLE\rlap{$^{\ddagger}$} & \CIRCLE & \CIRCLE & \CIRCLE\rlap{$^{\ddagger}$} & \CIRCLE & \CIRCLE & \CIRCLE & \CIRCLE & \CIRCLE & \CIRCLE & \CIRCLE & \CIRCLE & \CIRCLE & \CIRCLE \\
 & Postbox & GnuPG 2.5.24 / NSS 3.28.6 & \LEFTcircle & \CIRCLE\rlap{$^{\ddagger}$} & \Circle & \Circle & \Circle & \Circle & \Circle & \LEFTcircle & \LEFTcircle & \Circle & \Circle & \CIRCLE\rlap{$^{\ddagger}$} & \LEFTcircle & \LEFTcircle & \CIRCLE\rlap{$^{\ddagger}$} & \LEFTcircle & \CIRCLE & \CIRCLE & \CIRCLE & \Circle & \Circle & \Circle & \Circle & \Circle & \Circle \\
 & Thunderbird & RNP 0.18.1 / NSS 3.129 & \Circle & \Circle & \Circle & \Circle & \Circle & \Circle & \Circle & \Circle & \Circle & \Circle & \Circle & \Circle & \Circle & \Circle & \Circle & \Circle & \Circle & \Circle & \Circle & \LEFTcircle & \LEFTcircle & \LEFTcircle & \LEFTcircle & \LEFTcircle & \LEFTcircle \\
\addlinespace[2pt]
Linux & Claws Mail & GnuPG 2.5.24 / gpgsm 2.5.24 & \CIRCLE & \CIRCLE\rlap{$^{\ddagger}$} & \Circle & \Circle & \Circle & \Circle & \Circle & \CIRCLE & \CIRCLE & \Circle & \Circle & \CIRCLE\rlap{$^{\ddagger}$} & \CIRCLE & \CIRCLE & \CIRCLE\rlap{$^{\ddagger}$} & \CIRCLE & \CIRCLE & \CIRCLE & \CIRCLE & \Circle & \Circle & \Circle & \Circle & \Circle & \Circle \\
 & Thunderbird (Linux) & RNP 0.18.1 / NSS 3.129 & \Circle & \Circle & \Circle & \Circle & \Circle & \Circle & \Circle & \Circle & \Circle & \Circle & \Circle & \Circle & \Circle & \Circle & \Circle & \Circle & \Circle & \Circle & \Circle & \LEFTcircle & \LEFTcircle & \LEFTcircle & \LEFTcircle & \LEFTcircle & \LEFTcircle \\
 & Mutt & GnuPG 2.5.24 / OpenSSL 4.0.3 & \CIRCLE & \CIRCLE\rlap{$^{\ddagger}$} & \Circle & \Circle & \Circle & \Circle & \Circle & \CIRCLE & \CIRCLE & \Circle & \Circle & \CIRCLE\rlap{$^{\ddagger}$} & \CIRCLE & \CIRCLE & \CIRCLE\rlap{$^{\ddagger}$} & \CIRCLE & \CIRCLE & \CIRCLE & \CIRCLE & \CIRCLE & \CIRCLE & \CIRCLE & \CIRCLE & \CIRCLE & \CIRCLE \\
 & Evolution & GnuPG 2.5.24 / NSS 3.131 & \CIRCLE & \CIRCLE\rlap{$^{\ddagger}$} & \Circle & \Circle & \Circle & \Circle & \Circle & \CIRCLE & \CIRCLE & \Circle & \Circle & \CIRCLE\rlap{$^{\ddagger}$} & \CIRCLE & \CIRCLE & \CIRCLE\rlap{$^{\ddagger}$} & \CIRCLE & \CIRCLE & \CIRCLE & \CIRCLE & \LEFTcircle & \LEFTcircle & \LEFTcircle & \LEFTcircle & \LEFTcircle & \LEFTcircle \\
 & KMail & GnuPG 2.5.24 / gpgsm 2.5.24 & \CIRCLE & \CIRCLE\rlap{$^{\ddagger}$} & \Circle & \Circle & \Circle & \Circle & \Circle & \CIRCLE & \CIRCLE & \Circle & \Circle & \CIRCLE\rlap{$^{\ddagger}$} & \CIRCLE & \CIRCLE & \CIRCLE\rlap{$^{\ddagger}$} & \CIRCLE & \CIRCLE & \CIRCLE & \CIRCLE & \Circle & \Circle & \Circle & \Circle & \Circle & \Circle \\
\addlinespace[2pt]
Browser ext. & FlowCrypt & OpenPGP.js 6.3.2 / n/a& \Circle & \Circle & \Circle & \Circle & \Circle & \Circle & \Circle & \Circle & \Circle & \Circle & \Circle & \Circle & \Circle & \Circle & \Circle & \Circle & \Circle & \Circle & \Circle & n/a & n/a & n/a & n/a & n/a & n/a \\
 & Mailvelope & OpenPGP.js 5.11.3 / n/a& \Circle & \Circle & \Circle & \Circle & \Circle & \Circle & \Circle & \Circle & \Circle & \Circle & \Circle & \Circle & \Circle & \Circle & \Circle & \Circle & \Circle & \Circle & \Circle & n/a & n/a & n/a & n/a & n/a & n/a \\
\addlinespace[2pt]
Webmail & ProtonMail & OpenPGP.js 6.3.1 / n/a & \Circle & \Circle & \Circle & \Circle & \Circle & \Circle & \Circle & \Circle & \Circle & \Circle & \Circle & \Circle & \Circle & \Circle & \Circle & \Circle & \Circle & \Circle & \Circle & n/a & n/a & n/a & n/a & n/a & n/a \\
 & Roundcube & GnuPG 2.5.24 / n/a & \CIRCLE & \CIRCLE\rlap{$^{\ddagger}$} & \Circle & \Circle & \Circle & \Circle & \Circle & \CIRCLE & \CIRCLE & \Circle & \Circle & \CIRCLE\rlap{$^{\ddagger}$} & \CIRCLE & \CIRCLE & \CIRCLE\rlap{$^{\ddagger}$} & \CIRCLE & \CIRCLE & \CIRCLE & \CIRCLE & n/a & n/a & n/a & n/a & n/a & n/a \\
 & Horde & GnuPG 2.5.24 / PHP ext-openssl 4.0.3 & \LEFTcircle & \LEFTcircle & \Circle & \Circle & \Circle & \Circle & \Circle & \LEFTcircle & \LEFTcircle & \Circle & \Circle & \LEFTcircle & \LEFTcircle & \LEFTcircle & \LEFTcircle & \LEFTcircle & \LEFTcircle & \LEFTcircle & \LEFTcircle & \LEFTcircle & \LEFTcircle & \LEFTcircle & \LEFTcircle & \LEFTcircle & \LEFTcircle \\
 & Mailpile & GnuPG 2.5.24 / n/a& \CIRCLE & \CIRCLE\rlap{$^{\ddagger}$} & \Circle & \Circle & \Circle & \Circle & \Circle & \CIRCLE & \CIRCLE & \Circle & \Circle & \CIRCLE\rlap{$^{\ddagger}$} & \CIRCLE & \CIRCLE & \CIRCLE\rlap{$^{\ddagger}$} & \CIRCLE & \LEFTcircle & \LEFTcircle & \LEFTcircle & n/a & n/a & n/a & n/a & n/a & n/a \\
\addlinespace[2pt]
Android & FairEmail & PGPony 4.6.2 / BouncyCastle Java 1.77 & \CIRCLE & \LEFTcircle & \LEFTcircle & \LEFTcircle & \Circle & \Circle & \Circle & \LEFTcircle & \CIRCLE & \LEFTcircle & \Circle & \LEFTcircle & \LEFTcircle & \CIRCLE & \LEFTcircle & \LEFTcircle & \LEFTcircle & \LEFTcircle & \LEFTcircle & \Circle & \Circle & \Circle & \Circle & \Circle & \Circle \\
 & K-9 Mail & PGPony 4.6.2 / n/a & \CIRCLE & \LEFTcircle & \LEFTcircle & \LEFTcircle & \Circle & \Circle & \Circle & \LEFTcircle & \CIRCLE & \LEFTcircle & \Circle & \LEFTcircle & \LEFTcircle & \CIRCLE & \LEFTcircle & \LEFTcircle & \LEFTcircle & \LEFTcircle & \LEFTcircle & n/a & n/a & n/a & n/a & n/a & n/a \\
 & Thunderbird (Android) & PGPony 4.6.2 / n/a & \CIRCLE & \LEFTcircle & \LEFTcircle & \LEFTcircle & \Circle & \Circle & \Circle & \LEFTcircle & \CIRCLE & \LEFTcircle & \Circle & \LEFTcircle & \LEFTcircle & \CIRCLE & \LEFTcircle & \LEFTcircle & \LEFTcircle & \LEFTcircle & \LEFTcircle & n/a & n/a & n/a & n/a & n/a & n/a \\
 & Nine & n/a / BouncyCastle Java 1.57 & n/a & n/a & n/a & n/a & n/a & n/a & n/a & n/a & n/a & n/a & n/a & n/a & n/a & n/a & n/a & n/a & n/a & n/a & n/a & \Circle & \Circle & \Circle & \Circle & \Circle & \Circle \\
\addlinespace[2pt]
iOS & Canary Mail (iOS) & BouncyCastle Java 1.66 / n/a & \Circle & \Circle & \Circle & \Circle & \Circle & \Circle & \Circle & \Circle & \Circle & \Circle & \Circle & \Circle & \Circle & \Circle & \Circle & \Circle & \Circle & \Circle & \Circle & n/a & n/a & n/a & n/a & n/a & n/a \\
 & FlowCrypt (iOS) & OpenPGP.js 6.3.2 / n/a & \Circle & \Circle & \Circle & \Circle & \Circle & \Circle & \Circle & \Circle & \Circle & \Circle & \Circle & \Circle & \Circle & \Circle & \Circle & \Circle & \Circle & \Circle & \Circle & n/a & n/a & n/a & n/a & n/a & n/a \\
 & iOS Mail & iOS Keychain / apple.security.framework & n/a & n/a & n/a & n/a & n/a & n/a & n/a & n/a & n/a & n/a & n/a & n/a & n/a & n/a & n/a & n/a & n/a & n/a & n/a & \Circle & \Circle & \Circle & \Circle & \Circle & \Circle \\
\addlinespace[2pt]
macOS & Apple Mail & gmgmail~7.2 / apple.security.framework & \Circle & \Circle & \Circle & \Circle & \Circle & \Circle & \Circle & \Circle & \Circle & \Circle & \Circle & \Circle & \Circle & \Circle & \Circle & \Circle & \Circle & \Circle & \Circle & \Circle & \Circle & \Circle & \Circle & \Circle & \Circle \\
 & Canary Mail (macOS) & BouncyCastle Java 1.66 / n/a & \Circle & \Circle & \Circle & \Circle & \Circle & \Circle & \Circle & \Circle & \Circle & \Circle & \Circle & \Circle & \Circle & \Circle & \Circle & \Circle & \Circle & \Circle & \Circle & n/a & n/a & n/a & n/a & n/a & n/a \\
 & MailMate & GnuPG 2.5.24 / apple.security.framework& \CIRCLE & \CIRCLE\rlap{$^{\ddagger}$} & \Circle & \Circle & \Circle & \Circle & \Circle & \CIRCLE & \CIRCLE & \Circle & \Circle & \CIRCLE\rlap{$^{\ddagger}$} & \CIRCLE & \CIRCLE & \CIRCLE\rlap{$^{\ddagger}$} & \CIRCLE & \CIRCLE & \CIRCLE & \CIRCLE & \Circle & \Circle & \Circle & \Circle & \Circle & \Circle \\
 & Airmail & n/a / apple.security.framework& n/a & n/a & n/a & n/a & n/a & n/a & n/a & n/a & n/a & n/a & n/a & n/a & n/a & n/a & n/a & n/a & n/a & n/a & n/a & \Circle & \Circle & \Circle & \Circle & \Circle & \Circle \\
\bottomrule
\end{tabular}%
}
\end{table*}

\subsection{Provider and MX Observations}
\label{subsec:rq1_supporting}

\PP{Filtering-provider inference}
Following Rao~\etal~\cite{rao2024unfiltered}, the inferred Proofpoint and Google Workspace configuration of \texttt{zoom.us} combines MX records pointing to \texttt{pphosted.com}, a \texttt{250} reply from Google SMTP servers to an RCPT command for the \texttt{postmaster} address, and an SPF record containing \texttt{\_spf.google.com}.
This inference does not directly observe the internal forwarding connection.


\PP{Provider concentration and temporal attribution}
The submission and access (IMAP or POP3) populations contain 143,539 and 150,132 PQC-capable domains, respectively, as reported in the current analysis.
Migadu accounts for 54\% of the net relay increase from August~17 to August~31.
Strato accounts for 97\% and 83\% of the net submission and access increases from August~31 to September~14.
Among the 52,596 domains that host their own webmail, 3,261 gain hybrid support between August~31 and September~14 and 72 lose it (net $+$3,189).
Most of their webmail pages are served by cPanel (84\%) or Roundcube (10\%), whose TLS is terminated by the hosting server's own web server.
This suggests that independent hosting servers enabled hybrid key exchange.
Microsoft, Zoho, and Yandex show no support in any snapshot.
These observations cover four weeks and do not establish a long-term adoption trend.
The reported influence of Microsoft on intermediate delivery infrastructure provides additional context rather than a traffic-volume estimate from our measurements~\cite{li2025understanding}.

\PP{MX examples}
The domain \texttt{leboncoin.fr} has Tranco rank 1,207 and uses OVH backups.
The domain \texttt{hotstar.com} has rank 1,956 and uses a Proofpoint backup.
Both use \texttt{smtp.google.com} as a primary with hybrid support, while their backups lack it.
Across this category, Google commonly operates the primary, while Microsoft, Hostinger, Amazon, or Zoho operates a backup.
In the reverse category, Microsoft, Proofpoint, or Mimecast commonly operates the primary and Google a capable backup.
For example, \texttt{zendesk.com} (rank 331) lists Mimecast before Google.

\subsection{Message-Level Test Details}
\label{subsec:rq3_supporting}
This section describes the key corpus used in RQ3 in \S\ref{sec:rq3_result}.

\subsubsection{Key Corpus}
We introduce the key corpus for message-level encryption analysis.

\PP{Corpus design}
A PQC key can fail for independent reasons, so each key in \autoref{tab:keys}, except G1–G3, differs from at least one other corpus key in exactly
one factor. The baseline A1, the first RFC~9980 test vector, is the minimal conformant key: v6, the Ed25519 primary that RFC~9580 requires, and ML-KEM-768+X25519, the only KEM that RFC~9980 mandates~\cite{rfc9980}. 
A2–A7 are the remaining RFC 9980 test vectors: A2 moves A1 to v4, A3, A5, and A6 replace A1's primary with ML-DSA or SLH-DSA, and A4 and A7 do the same for N2, which carries an ML-KEM-1024+X448 subkey.
The other keys cover combinations for which RFC~9980 provides no test vector. Starting from A1, (i)~N1 changes only the key version (v5), N2 changes only the subkey (ML-KEM-1024+X448), and N3 and N4 change both; the latter two are combinations that RFC~9980 prohibits and serve as negative controls. (ii)~R-v4/v5/v6 and L-v4/v5 replace the primary with RSA-3072 and legacy EdDSA, respectively, to test whether other primary algorithms work across key versions. (iii)~G1--G3 are generated by GnuPG, which follows LibrePGP rather than RFC~9980, and test interoperability with the dominant backend. The S/MIME certificates O1--O6 cover three ML-KEM levels in standard and legacy PKCS\#12 packaging.

\subsubsection{Detailed Per-Client Support Matrix}
\label{appx:subsubsec:rq3_encryption_Decrpyion}
Table~\ref{tab:matrix} shows the 27 client–platform configurations for importing and encryption.

\PP{Negative Controls}
N3 and N4 pair an Ed25519 primary key with an ML-KEM-1024+X448 subkey in v5 and v4 formats, respectively, combinations that RFC~9980 prohibits.
Rejecting them at import is therefore the correct outcome.
All GnuPG-based configurations reject both keys.
The three PGPony-based configurations reject N4 and import N3, indicating that their import stage performs no conformance check against RFC~9980.
Encryption to N3 then fails, as for every other v5 key, because PGPony accepts only v6 keys at encryption time (\S\ref{subsec:rq3_exchange}).
The failure thus stems from the key version rather than from the prohibited combination, and no message is sent under the nonconforming key.
These results motivate validating imported keys against RFC~9980 at import time.

\PP{GnuPG PKESK Encoding Flaw}
GnuPG~2.5.24 writes the 20-octet v4 fingerprint in the PKESK packet followed by twelve zero octets, consistent with a field sized for 32-octet v6 fingerprints.
The recipient therefore reads a zero where the algorithm identifier should be.
Removing the twelve octets from the generated messages restores decryption.

\PP{Postbox}
The Enigmail~3.1.8 layer of Postbox derives the key ID from the last 16 hex digits of the fingerprint, which is correct only for v4 keys, so GnuPG receives a nonexistent recipient for v5 and v6 keys.
Its v4 keys behave as in the other GnuPG-based clients ($\ddagger$).
G1--G3 succeed because their v4 primary key satisfies the key-ID rule and their v5 subkey avoids the PKESK flaw.

\PP{Horde and Mailpile}
Horde invokes GnuPG with the obsolete \texttt{--secret-keyring} option and treats the resulting warning as an error, so every exchange fails regardless of the key.
This failure is unrelated to PQC.
Mailpile indexes keys by fingerprint and does not handle keys whose primary and subkey fingerprints differ in length.
G1--G3 combine a 20-octet v4 primary fingerprint with a 32-octet v5 subkey fingerprint, so Mailpile loses their key records after import.

\PP{PGPony}
PGPony imports keys with ML-DSA primaries by falling back from bc-java to its own implementation, which alone supports ML-DSA.
Encryption and decryption go only through bc-java, so these keys fail after import.

\PP{S/MIME Integration}
Windows CNG has provided \texttt{KEMRecipient\allowbreak Info} since the November 2025 update, and OpenSSL since version~3.6.
NSS (Thunderbird, Evolution) accepts only RSA and EC public keys as recipients.
eM Client imports the certificates because .NET's X.509 layer delegates to Windows CNG, and builds the envelope with .NET's System.\allowbreak Security.\allowbreak Cryptography.\allowbreak Pkcs~9.0.1, which does not implement \texttt{KEMRecipientInfo}.
Horde uses the same OpenSSL version as Mutt and builds the envelope through OpenSSL's legacy PKCS\#7 API, which lacks \texttt{KEMRecipientInfo}, whereas Mutt uses the CMS API.

\end{document}